\documentclass[aps,prd,preprint,a4paper,portrait,preprintnumbers,nofootinbib,tightenlines,unsortedaddress]{revtex4-1}
\usepackage{graphicx}
\usepackage{amsmath}
\usepackage{psfrag}
\usepackage{slashed}
\usepackage{hyperref}

\graphicspath{{figures/}}

\def\la{\langle}
\def\ra{\rangle}

\def\A#1#2{\la#1#2\ra}
\def\B#1#2{[#1#2]}
\def\AB#1#2#3{\la#1|#2|#3]}

\def\AA#1#2#3{\la#1|#2|#3\ra}

\DeclareMathOperator{\tr}{\rm tr}
\def\mc{\mathcal}

\def\Frac[#1,#2]{\frac{#1}{#2}} 
\def\nn{\nonumber}

\def\eps{\epsilon}
\def\fl#1{#1^\flat}

\def\kf#1{K_{#1}^\flat}
\def\kfm#1{K_{#1}^{\flat,\mu}}
\def\tb{{\bar{t}}}
\def\bb{{\bar{b}}}
\def\qb{{\bar{q}}}
\def\ib{{\bar{\imath}}}

\begin{document}

\title{One-Loop Helicity Amplitudes for $t\tb$ Production\\ at Hadron Colliders.}
\date{\today}

\author{Simon Badger}
\affiliation{The Niels Bohr International Academy and Discovery Center,\\
  The Niels Bohr Institute, Blegdamsvej 17,\\
  DK-2100 Copenhagen, Denmark}
\email{simon.badger@nbi.dk}
\author{Ralf Sattler}
\affiliation{Institut f\"ur Physik, Humboldt-Universit\"at zu Berlin, Newtonstra\ss{}e 15,\\
  D-12489 Berlin, Germany}
\affiliation{Deutsches Elektronen-Synchrotron DESY, Platanenallee 6,\\ D-15738 Zeuthen, Germany}
\email{ralf.sattler@physik.hu-berlin.de}
\author{Valery Yundin}
\affiliation{Institute of Physics, University of Silesia, Uniwersytecka 4,\\ PL-40007 Katowice, Poland}
\affiliation{Deutsches Elektronen-Synchrotron DESY, Platanenallee 6,\\ D-15738 Zeuthen, Germany}
\email{valery.yundin@desy.de}
\author{ }
\affiliation{ }
\begin{abstract}
We present compact analytic expressions for all one-loop helicity amplitudes
contributing to~$t\bar{t}$ production at hadron colliders. Using recently developed generalised 
unitarity methods and a traditional Feynman based approach we produce a fast and flexible
implementation.
\end{abstract}

\preprint{LPN-11-06}
\preprint{DESY-11-012}
\preprint{HU-EP-11/06}
\preprint{SFB/CPP-11-04}

%\keywords{QCD, Hadron Colliders, Higher-Order Corrections}

\maketitle

\section{Introduction}

The large production rate of top quarks at the Large Hadron Collider (LHC) 
\cite{Bernreuther:2008ju}, and the high numbers of leptonic decays, make it a key signal for the 
study of precision physics in the Standard Model (SM). The first theoretical predictions for the  
unpolarised cross section at next-to-leading order (NLO) have been known for over twenty years
\cite{Nason:1987xz,Nason:1989zy,Beenakker:ttgg,Frixione:1995fj}. However, recent years have seen 
substantial progress in NLO predictions for heavy quark production and associated processes.  
Spin correlations and on-shell decays of the top quarks at NLO 
\cite{Bernreuther:ttggspin,Melnikov:2009dn} as well as electro-weak corrections 
\cite{Beenakker:1993yr,Bernreuther:2005is,Kuhn:2005it,Moretti:2006nf,Bernreuther:2006vg,Kuhn:2006vh,Hollik:2007sw} 
are now well understood. One-loop amplitudes to higher order in the dimensional regularisation 
parameter, which form part of the NNLO prediction, have also been computed 
\cite{Korner:ttgge2,Anastasiou:ttgge2}. Predictions for 
processes with heavy quark production in association with
other QCD particles have also been made possible thanks to impressive calculations of $pp\to t\tb+j$%
~\cite{Dittmaier:2007wz,Dittmaier:2008uj,Ellis:ttggg,Melnikov:2010iu}, $pp\to t\tb 
b\bb$~\cite{Bredenstein:2008zb,Bredenstein:2009aj,Bredenstein:2010rs,Bevilacqua:2009zn} and $pp\to 
t\tb+2j$~\cite{Bevilacqua:2010ve}.
Very recently, full off-shell effects of the top decays have been computed 
\cite{Denner:2010jp,Bevilacqua:2010qb}.

The last few years have seen a rapid development in NLO techniques allowing the computation of new 
multi-leg amplitudes and cross-sections. On-shell techniques, which began with the development of the unitarity method 
\cite{Bern:uni1,Bern:uni2}, are quickly becoming an industry standard tool and, through working with 
gauge invariant building blocks, lead to extremely fast numerical evaluation for the virtual 
corrections to NLO processes \cite{Berger:2008sj,Ellis:numgu,Giele:ddnumgu,Ossola:OPP,Giele:2008bc}. 
Some of these methods have also been developed into public codes \cite{Mastrolia:2010nb,Ossola:CutTools,Badger:2010nx}.  
Together with advanced Feynman diagram based techniques many phenomenological predictions have been possible for $2\to4$ processes 
\cite{Bredenstein:2008zb,Bredenstein:2009aj,Bredenstein:2010rs,Ellis:2009zw,KeithEllis:2009bu,Berger:2009zg,Berger:2009ep,Berger:2010vm,Melia:2010bm,Bevilacqua:2009zn,Bevilacqua:2010ve,Bevilacqua:2010qb,Denner:2010jp}. 
Last year also saw the first evaluations of $2\to5$ cross-sections with computations of $pp\to W+4j$ 
\cite{Berger:2010zx} and $e^+e^-\to 5j$ \cite{Frederix:2010ne}.
The extension of analytic unitarity techniques to deal with massive particles motivates the 
revisiting of the well known process of heavy quark
production in hadron collisions. In this paper we construct compact analytic expressions for the 
virtual helicity amplitudes. These amplitudes should lead to a flexible evaluation of the NLO 
cross-section including both spin correlations and decays in the narrow width approximation. We 
demonstrate that the amplitudes presented here evaluate a factor of $\sim10$ times faster 
than the analytic results of ref. \cite{Korner:ttgg} implemented in {\tt MCFM} \cite{mcfm}. Full 
analytic computations also offer the possibility of investigating new structures in gauge theory 
amplitudes. In our particular example we find new simplicity in the sub-leading colour 
contributions to the one-loop amplitudes.

Our paper is organised as follows. We first review the decomposition of the full one-loop amplitude 
into colour ordered and primitive amplitudes which form the basic building blocks of our 
computation. We then give our notation for the spinor-helicity formalism used for the computation of 
helicity amplitudes with massive particles. In section \ref{sec:methods} we outline the unitarity 
and Feynman based methods employed to arrive at the compact expression. We give details of the pole 
structures and renormalisation procedure in sections \ref{sec:IR} and \ref{sec:renorm} before 
presenting the complete set of independent helicity amplitudes in \ref{sec:treeamps} and 
\ref{sec:1lamps}. Some details of the numerical implementation are given in section \ref{sec:num} 
before we reach our conclusions. An appendix listing the complete set of tree-level amplitudes 
is included for completeness.

\section{Colour Ordering and Primitive Amplitudes \label{sec:colour}}

We follow the colour decomposition into primitive amplitudes as described by Bern, Dixon and
Kosower \cite{Bern:1lqqggg}. The case of massive quarks is identical to that of the massless case. 
Although we will talk of the processes $gg\to t\tb$ and $q\qb\to t\tb$ throughout the paper, it should be noted that 
the amplitudes are computed with all particles considered to be outgoing.

\subsection{The $gg\to t\tb$ Channel}

Firstly the tree-level amplitude can be written,
\begin{align}
   \mc A^{(0)}_4(1_t,2,3,4_\tb) =& g_s^2 \sum_{P(2,3)}
   (T^{a_2}T^{a_3})_{i_1}^{\ib_4} A^{(0)}_{4}(1_t,2,3,4_\tb),
\end{align}
while the one-loop amplitude reads,
\begin{align}
  \mc A^{(1)}_4(1_t,2,3,4_\tb) =& g_s^2c_\Gamma \bigg (\sum_{P(2,3)} N_c 
  (T^{a_2}T^{a_3})_{i_1}^{\ib_4} A^{(1)}_{4;1}(1_t,2,3,4_\tb)\nonumber\\&
  +\tr(T^{a_2}T^{a_3})\delta_{i_1}^{\ib_4} A^{(1)}_{4;3}(1_t,4_\tb;2,3)
  \bigg),
\end{align}
where, in $D=4-2\eps$ dimensions, $c_\Gamma$ is defined as,
\begin{equation}
  c_\Gamma = \frac{\Gamma(1+\eps)\Gamma^2(1-\eps)}{(4\pi)^{2-\eps}\Gamma(1-2\eps)}.
  \label{eq:cGamma}
\end{equation}
The one-loop amplitudes can be further decomposed into gauge invariant primitive amplitudes,
\begin{align}
	A^{(1)}_{4;1}(1_t,2,3,4_\tb) &= 
	A_4^{[L]}(1_t,2,3,4_\tb)-\frac{1}{N_c^2}A_4^{[R]}(1_t,2,3,4_\tb)\nonumber\\&
	-\frac{N_f}{N_c} A_4^{[f]}(1_t,2,3,4_\tb)-\frac{N_H}{N_c} A_4^{[H]}(1_t,2,3,4_\tb),\\
	A^{(1)}_{4;3}(1_t,4_\tb;2,3) &= \sum_{P(2,3)} \bigg\{
		A_4^{[L]}(1_t,2,3,4_\tb)
		+A_4^{[L]}(1_t,2,4_\tb,3)
		+A_4^{[R]}(1_t,2,3,4_\tb)\bigg\},
\end{align}
where $N_c$ is the number of colours, $N_f$ and $N_H$ are the number of light and heavy flavours 
respectively. Explicitly the permutation set is $P(2,3)=\{(2,3),(3,2)\}$.

Performing the colour summations for the squared tree level amplitudes yields,
\begin{align}
	\sum_{\rm col} |A_4^{(0)}|^2 &= 
	g_s^4 N_c(N_c^2-1) \, \sum_{P(2,3)} |A_{4}^{(0)}(1_t,2,3,4_\tb)|^2 -\frac{N_c^2-1}{N_c}\bigg|\sum_{P(2,3)}
	A_{4}^{(0)}(1_t,2,3,4_\tb) \bigg|^2\\
	&=
	g_s^4 N_c(N_c^2-1) \, \sum_{P(2,3)} |A_{4}^{(0)}(1_t,2,3,4_\tb)|^2-\frac{N_c^2-1}{N_c}|A_{4}^{(0)}(1_t,2_\gamma,3_\gamma,4_\tb) |^2.
\end{align}
For the interference between the tree and one-loop amplitudes we find
\begin{multline}
  \sum_{\rm col} \mc A_4^{(1)}[\mc A_4^{(0)}]^* =
  g_s^6 c_\Gamma N_c^2 (N_c^2-1) \, \sum_{P(2,3)} 
  A_{4;1}^{(1)}(1_t,2,3,4_\tb)[A_{4}^{(0)}(1_t,2,3,4_\tb)]^* \\
  + (N_c^2-1) \left(A_{4;3}^{(1)}(1_t,4_\tb;2,3)-A_{4;1}^{(1)}(1_t,2,3,4_\tb) \right. \\ 
  \left. - A_{4;1}^{(1)}(1_t,3,2,4_\tb) \right) \, [A_{4}^{(0)}(1_t,2_\gamma,3_\gamma,4_\tb)]^* \, ,
  \label{eq:1lggCHsum}
\end{multline}

\subsection{The $q\qb\to t\tb $ Channel}
This time the tree-level amplitude is  
\begin{equation}
 \mathcal{A}_4^{(0)}\left( 1_t , 2_\qb, 3_q , 4_\tb,\right) = g_s^2 \left( \delta_{i_1}^{\ib_2} 
 \delta_{i_3}^{\ib_4} - \frac{1}{N_c} \delta_{i_1}^{\ib_4} \delta_{i_3}^{\ib_2} \right) 
 A_4^{(0)}\left(1_t,2_\qb, 3_q, 4_\tb \right) \, ,
\end{equation}
while for the decomposition of the one-loop amplitude we find
\begin{equation} 
    \mathcal{A}_4^{(1)}\left( 1_t, 2_\qb, 3_q, 4_\tb \right) = 
    g_s^4 c_\Gamma \left(N_c \, \delta_{i_1}^{\ib_2} \delta_{i_3}^{\ib_4} \, A_{4;1}^{(1)}\left( 1_t, 2_\qb, 3_q, 4_\tb \right) 
    + \delta_{i_1}^{\ib_4} \delta_{i_3}^{\ib_2} \, A_{4;2}^{(1)}\left( 1_t, 2_\qb, 3_q, 4_\tb 
    \right) \right)\, ,
\end{equation}
with the primitives
\begin{align}
A_{4;1}^{(1)}&\left( 1_t, 2_\qb, 3_q, 4_\tb \right)  = 
A_4^{[lc]}\left( 1_t, 2_\qb, 3_q, 4_\tb \right) 
-\frac{N_f}{N_c} A_4^{[f]}\left( 1_t, 2_\qb, 3_q, 4_\tb \right)
-\frac{N_H}{N_c} A_4^{[H]}\left( 1_t, 2_\qb, 3_q, 4_\tb \right)
\nonumber\\&
- \frac{2}{N_c^2} \left( A_4^{[lc]}\left(  1_t, 2_\qb, 3_q, 4_\tb \right) 
+ A_4^{[lc]}\left( 1_t, 3_q, 2_\qb, 4_\tb \right) \right)
- \frac{1}{N_c^2} A_4^{[slc]}\left( 1_t, 2_\qb, 3_q, 4_\tb  \right)
\end{align}
and 
\begin{align}
A_{4;2}^{(1)}&\left( 1_t, 2_\qb, 3_q, 4_\tb \right) = 
A_4^{[lc]}\left(  1_t, 3_q, 2_\qb, 4_\tb  \right) 
-\frac{N_f}{N_c} A_4^{[f]}\left( 1_t, 3_q, 2_\qb, 4_\tb \right)
-\frac{N_H}{N_c} A_4^{[H]}\left( 1_t, 3_q, 2_\qb, 4_\tb \right)
\nonumber\\&
+ \frac{1}{N_c^2} \left( A_4^{[lc]}\left(  1_t, 3_q, 2_\qb, 4_\tb \right) + A_4^{[lc]}\left( 1_t, 2_\qb, 
3_q, 4_\tb  \right) \right)
+ \frac{1}{N_c^2} A_4^{[slc]}\left( 1_t, 3_q, 2_\qb, 4_\tb  \right) \,.
\end{align}
Evaluating the colour ordered squared tree-level amplitude yields then
\begin{equation}
 \sum_{\rm col} \, \lvert \mathcal{A}_4^{(0)} \rvert = g_s^4 (N_c^2 - 1) \, \lvert 
 A_{4}^{(0)}(1_t,2_\qb,3_q,4_\tb) \rvert^2 \, ,
\end{equation}
where as the interference term becomes
\begin{equation}
  \sum_{\rm col} \mathcal{A}_4^{(1)} [\mathcal{A}_4^{(0)}]^* = g_s^6 c_\Gamma N_c (N_c^2 - 1) \, 
  A_{4;1}^{(1)}(1_t,2_\qb,3_q,4_\tb) \, 
  [A_4^{(0)}(1_t,2_\qb,3_q,4_\tb)]^* \, .	
\end{equation}

\section{Spinor-helicity formalism \label{sec:spinors}}

For massless particles it is possible to completely decompose all momenta into a basis of
two component Weyl spinors since,
\begin{equation}
  p^\mu = \tfrac{1}{2}\AB{p}{\gamma^\mu}{p}.
  \label{eq:mdecomp}
\end{equation}
The polarisation vectors and fermion wave functions then fit easily into a helicity basis:
\begin{align}
	u_+(p) &= |p\ra, & u_-(p) &= |p], \\
	\eps_+^\mu(p,\xi) &= \frac{\AB{\xi}{\gamma^\mu}{p}}{\sqrt{2}\A{\xi}{p}}, &
	\eps_-^\mu(p,\xi) &= \frac{\AB{p}{\gamma^\mu}{\xi}}{\sqrt{2}\B{p}{\xi}}.
\end{align}
The situation for massive momenta is a little more complicated since \eqref{eq:mdecomp} is no longer
valid and the definition of unique helicity state is no longer possible as one can always find a
boost, through the rest frame, such that the helicity state is flipped. However, given a massive
momentum $P$, one can construct a helicity basis by defining a specific reference frame with respect
to an arbitrary massless vector $\eta$ \cite{Kleiss:massspin},
\begin{equation}
  P^\mu = \alpha \fl{P} + \beta \eta_P
\end{equation}
where $\alpha\beta = \tfrac{m^2}{\A{\fl P}{\eta_P}\B{\eta_P}{\fl P}}$.
The $u$ and $v$ spinors can then be defined by:
\begin{align}
  u_\pm(P,m;\fl P,\eta_P) &= \frac{(\slashed{P}+m)|\eta_P\pm\ra}{\sqrt{\alpha}\A{\fl P \mp|}{\eta_P\pm}},
  &
  \bar u_\pm(P,m;\fl P,\eta_P) &= \frac{\la\eta_P\!\mp\!|(\slashed{P}+m)}{\sqrt{\alpha}\A{\eta_P\mp|}{\fl P \pm}}, 
  \label{eq:Usp_m}\\
  v_\pm(P,m;\fl P,\eta_P) &= \frac{(\slashed{P}-m)|\eta_P\pm\ra}{\sqrt{\alpha}\A{\fl P \mp|}{\eta_P\pm}},
  &
  \bar v_\pm(P,m;\fl P,\eta_P) &= \frac{\la\eta_P\!\mp\!|(\slashed{P}-m)}{\sqrt{\alpha}\A{\eta_P\mp|}{\fl P \pm}},
  \label{eq:Vsp_m}
\end{align}
in the following we will set $\alpha=1$, $\beta=\tfrac{m^2}{2\fl P\cdot\eta_P}$ which corresponds to the choice of
 reference~\cite{Schwinn:heavyQ} although a choice of,
\begin{equation}
	\alpha = \frac{1}{2}\left(1+\sqrt{1-\tfrac{4m^2}{2\fl P\cdot\eta_P}}\right),
 \qquad
 	\beta = \frac{1}{2}\left(1-\sqrt{1-\tfrac{4m^2}{2\fl P\cdot\eta_P}}\right),
\end{equation}
corresponds to the basis used in reference \cite{Rodrigo:heavyQ,Hagiwara:massspin}
which also has a convenient massless limit. A specific choice of the reference vector that allows us 
to match the four component representation of ref. \cite{Ellis:ttggg} is presented in Appendix 
\ref{App:EGKMref}. Keeping the reference vectors $\eta$ arbitrary allows us to relate the heavy quark helicity states:
\begin{align}
  u_-(P,m;\fl P,\eta_P) &= -\frac{\A{\fl P}{\eta_P}}{m} u_+(P,m;\eta_P,\fl P), \\
  v_-(P,m;\fl P,\eta_P) &= \frac{\A{\fl P}{\eta_P}}{m} v_+(P,m;\eta_P,\fl P).
  \label{eq:helswapid}
\end{align}
The spinor-helicity formalism is well suited for numerical evaluation and had been implemented for
mathematica within the {\tt S@M} package \cite{Maitre:SaM}.

\section{Methods \label{sec:methods}}

\begin{figure}[h]
  \begin{center}
    \includegraphics[width=12cm,trim=0pt 16pt 0pt 4pt]{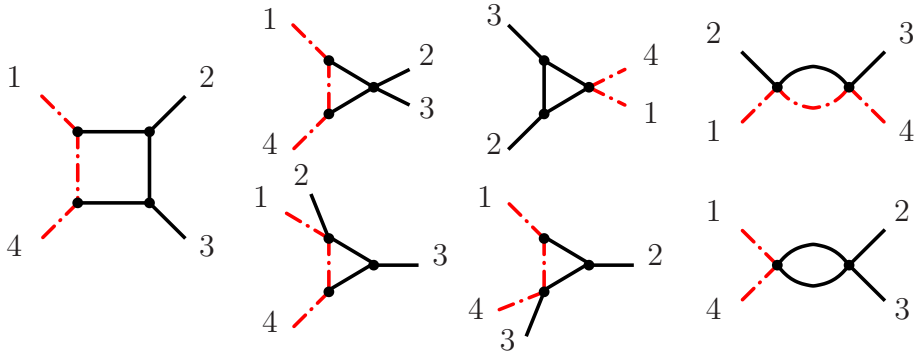}
  \end{center}
  \caption{The 7 cut diagrams contributing to the left-moving primitive amplitude in $gg\to t\tb$. Red~dotted lines represent massive fermions, plain 
  lines represent gluons.}
  \label{fig:intbasisL}
\end{figure}

\begin{figure}[h]
  \begin{center}
    \includegraphics[width=12cm,trim=0pt 16pt 0pt 8pt]{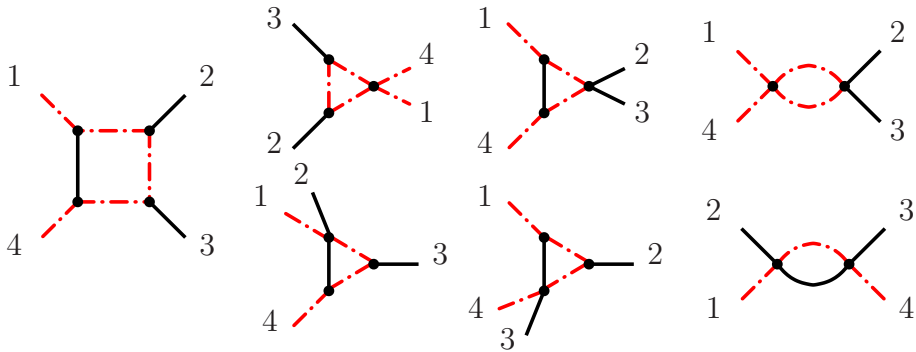}
  \end{center}
  \caption{The 7 cut diagrams contributing to the right-moving primitive amplitude in $gg\to t\tb$. Red~dotted lines represent massive fermions, plain
  lines represent gluons.}
  \label{fig:intbasisR}
\end{figure}

\begin{figure}[h]
  \begin{center}
    \includegraphics[width=12cm,trim=0pt 18pt 0pt 4pt]{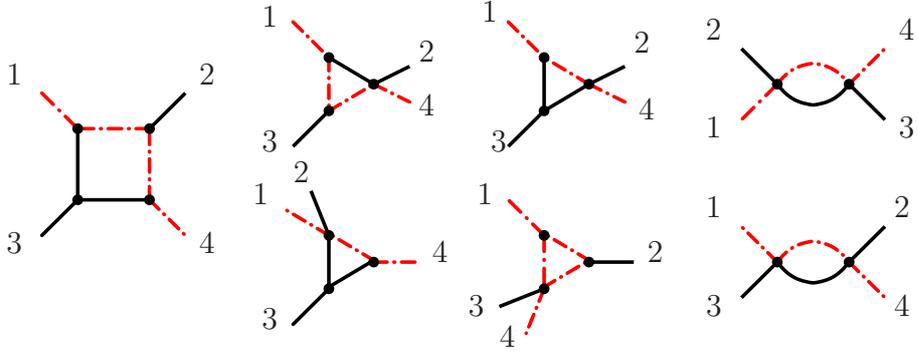}
  \end{center}
  \caption{The 7 cut diagrams contributing to the sub-leading colour primitive amplitude in $gg\to t\tb$. Red~dotted lines represent massive fermions, plain
  lines represent gluons. %We find only the box and triangle diagrams contribute to the full 
  amplitude.}
  \label{fig:intbasisLsl}
\end{figure}

\begin{figure}[h]
  \begin{center}
    \includegraphics[width=12cm,trim=0pt 18pt 0pt 16pt]{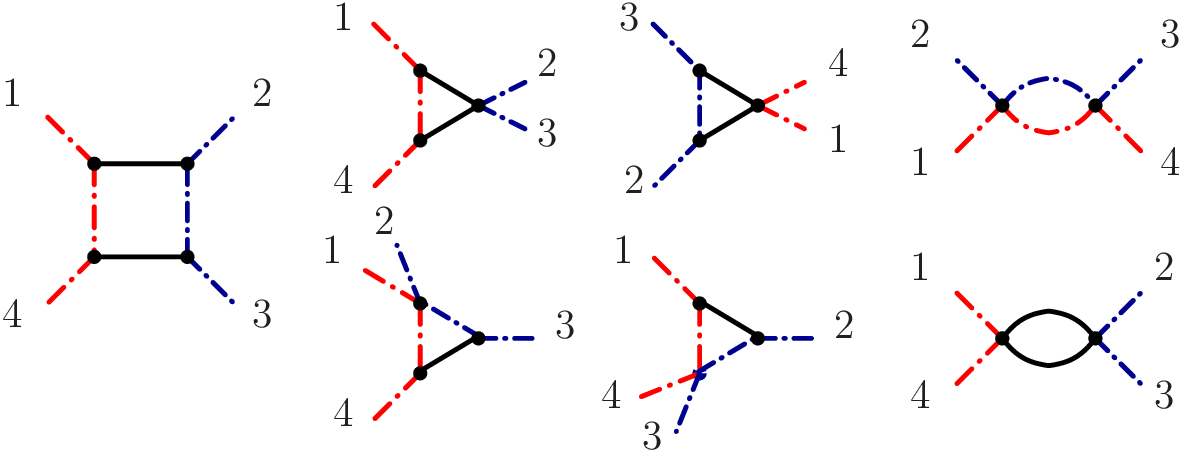}
  \end{center}
  \caption{The 7 cut diagrams contributing to the leading colour primitive amplitude in $q\qb\to t\tb$. Red~dotted lines represent massive fermions, plain
  lines represent gluons and blue dotted lines represent massless fermions.}
  \label{fig:intbasislc}
\end{figure}

\begin{figure}[h]
  \begin{center}
    \includegraphics[width=6cm,trim=0pt 12pt 0pt 16pt]{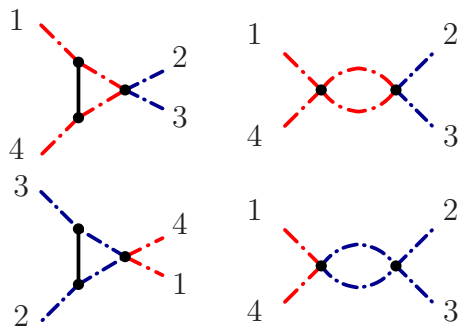}
  \end{center}
  \caption{The 4 cut diagrams contributing to the sub-leading colour primitive amplitude in $q\qb\to t\tb$. Red~dotted lines represent massive fermions, plain
  lines represent gluons and blue dotted lines represent massless fermions.}
  \label{fig:intbasisslc}
\end{figure}

Each of our primitive one-loop amplitudes can be written using the standard basis of scalar integral 
functions and a rational term up to higher order terms in the dimensional regularisation parameter $\eps$,
\begin{align}
  A^{[X]}_{4}&(1,2,3,4) = C^{[X]}_{4;1|2|3|4} \, I^{[X]}_{4;1|2|3|4} \nonumber\\&
  + \sum_{i=1}^{2}\sum_{j=i+1}^{3}\sum_{k=j+1}^{i-1}
  C^{[X]}_{3;i\dots j-1|j\dots k-1|k\dots i-1} \, I^{[X]}_{3;i\dots j-1|j\dots k-1|k\dots 
  i-1}\nonumber\\&
  + \sum_{i=1}^{3}\sum_{j=i+1}^{i-1} C^{[X]}_{2;i\dots j-1|j\dots i-1} \, I^{[X]}_{2;i\dots 
  j-1|j\dots i-1} + C^{[X]}_{1} I^{[X]}_{1} + R^{[X]} + \mathcal{O}(\eps).
  \label{eq:scalar}
\end{align}
The indices $\{i,j,k,l\}$ denote the momenta at the external legs of the integrals and the sums run
over all cyclic partitions. The dependence of the coefficients and the
rational term on the helicity states of the external particles has been suppressed above. Figures 
\ref{fig:intbasisL}-\ref{fig:intbasisslc} show the explicit basis integrals for the five independent 
primitive amplitudes. Heavy and light flavour fermion loop contributions have also been included.

The computation was performed in two distinct parts. The first used generalised unitarity to compute 
the compact expressions for the coefficients of the scalar integrals. A Feynman diagram based 
approach was then taken to obtain compact forms for the remaining tadpoles and rational terms.

The computation was performed in the Four-Dimensional-Helicity (FDH) scheme and mass 
renormalisation was performed using the on-shell scheme.

\subsection{Generalised Unitarity}

The use of the generalised unitarity 
\cite{Anastasiou:ddcuts,Anastasiou:gumaster,Britto:genu,Forde:intcoeffs,Giele:ddnumgu,Mastrolia:2009dr}, 
extended to massive propagators \cite{Bern:massuni,Britto:gumassive,Kilgore:massuni,Ellis:ttggg}, is the primary reason we are able 
to obtain compact representations for the one-loop helicity amplitudes.
The analytic techniques we have employed are by now well covered in the 
existing literature and have been used extensively in recent analytic computations of $pp\to 
H+2j$~\cite{Dixon:2009uk,Badger:2009hw,Badger:2009vh} and $pp\to Wb\bb$~\cite{Badger:2010mg}
production. We refer the reader to recent reviews on the topic for further details 
\cite{Britto:2010xq,Berger:2009zb}. However, since analytic 
computations for massive amplitudes are covered to a lesser extent, we outline some of the 
techniques specific to our process.

$D$-dimensional cutting procedures \cite{Bern:massuni,Giele:ddnumgu,Britto:2008vq,Badger:2008cm} could also be applied to the computation of rational terms in 
massive amplitudes as described in ref.~\cite{Ellis:ttggg}. However, as we discuss briefly in 
section \ref{sec:tad}, the procedure can lead to large intermediate expressions when followed 
analytically. On-shell recursion relations offer an attractive alternative to obtain compact 
expressions directly \cite{Bern:1lrec,Bern:2005ji,Bern:2005cq,Berger:2006ci}. At the present time 
such techniques have yet to be extended to the massive case. A Feynman diagram approach therefore 
offers a simple way to obtain compact analytic expressions for our process.

\subsubsection{Three-Mass Triangle Coefficients}

In this section we give an explicit example of the computation for the three-(external)-mass triangle coefficient $C_{3;1|23|4}$. This gives a good example of a 
computation that is specific to the massive case. We consider the $++-+$ helicity configuration in 
the $gg\to t\tb$ since the $++++$ configuration is 
zero. Following Forde's method \cite{Forde:intcoeffs}, the triple cut can be parametrised by a 
one-dimensional complex contour integral over $t$. The on-shell constraints for the loop momentum 
are solved in general leaving the integrand as a rational function of $t$. The scalar triangle 
coefficient is then given as the boundary value of this triple cut integrand:
\begin{align}
  &C^{[L]}_{3;4|1|23}(1_t^+,2^+,3^-,4_\tb^+) = -\frac{1}{2}\sum_{\gamma_\pm}{\rm inf}_t [ \tilde{C}^{[L]}_{3;4|1|23}(1_t^+,2^+,3^-,4_\tb^+)  ]\bigg|_{t^0},
  \\& 
  \tilde{C}^{[L]}_{3;4|1|23}(1_t^+,2^+,3^-,4_\tb^+) = 
  \nonumber\\& 
  \sum_{h_1=\pm}
  \sum_{h_2=\pm}
  \sum_{h_3=\pm}
  A^{(0)}(-l_3^{-h_3},4_\tb^+,l_{1;t}^{h_1})
  A^{(0)}(-l_{1;\tb}^{-h_1},1_t^+,l_2^{h_2})
  A^{(0)}(-l_2^{-h_2},2^+,3^-,l_3^{h_3}),
  \label{eq:3cutA}
\end{align}
where ${\rm inf}_t$ is computed by taking a Taylor expansion around $t=\infty$. The complex parameter $t$ appears in the parametrisation of the 
(on-shell) loop momentum. We construct a spinor basis for the complex loop momentum using two massless vectors $\kf1$ and 
$\kf2$. In our case the massless vectors can be constructed from $p_4$ and $p_1$ such that,
\begin{align}
  \kfm1 &= \frac{\gamma(\gamma p_4^\mu - m^2 p_1^\mu)}{\gamma^2-m^4},\\
  \kfm2 &= \frac{\gamma(\gamma p_1^\mu - m^2 p_4^\mu)}{\gamma^2-m^4},
\end{align}
where $\gamma_\pm = p_1\cdot p_4 \pm \sqrt{(p_1\cdot p_4)^2-m^4}$. The on-shell loop momentum $l_1$ can then be written:
\begin{align}
  l_1^\mu = a \left( \kfm1 - \kfm2 \right)
  + \frac{t}{2} \AB{\kf1}{\gamma^\mu}{\kf2}
  - \frac{a^2\gamma + m^2 }{2\gamma t} \AB{\kf2}{\gamma^\mu}{\kf1},
  \label{eq:lparam}
\end{align}
where
\begin{align}
  a = \frac{m^2}{\gamma-m^2}.
\end{align}
The procedure from this point is rather straightforward. For illustrative purposes we find it convenient to expand the helicity sum
 in eq.~(\ref{eq:3cutA}) and explicitly remove dependence of the internal reference vector of the massive spinor. This results in considerably shorter 
expressions for the integrand though in general we find that specific choices of this vector yield comparable sized expressions after expansion. 
Inserting the relevant tree amplitudes from Appendix~\ref{App:trees} we find (the reference vectors for the internal gluons were chosen as 
$\xi_{l_2}^\mu=\xi_{l_3}^\mu =  \tfrac{1}{2}\AB{\kf2}{\gamma^\mu}{\kf1}$):
\begin{align}
  -i\tilde{C}^{[L]}_{3;4|1|23}&(1_t^+,2^+,3^-,4_\tb^+) =
  \nonumber\\&
  \frac{ m\la\kf2|1|l_2]\left(\la \eta_4\eta_1\ra\la l_3|4|\kf1]+\la l_3\eta_1\ra\la l_3\eta_4\ra[\kf1l_3]\right)\la l_23\ra^4}
  {\la\eta_4\fl4\ra\la\kf2l_2\ra\la l_2l_3\ra\la l_22\ra\la l_33\ra\la\fl1\eta_1\ra\la23\ra[\kf1l_3]}
  \nonumber\\&
  +\frac{ m\la l_33\ra^3\la\kf2|4|l_3]\left(\la\eta_4\eta_1\ra\la l_2|1|\kf1]+\la l_2\eta_1\ra\la l_2\eta_4\ra[\kf1l_2]\right)}
  {\la\eta_4\fl4\ra\la\kf2l_3\ra\la l_2l_3\ra\la l_22\ra\la\fl1\eta_1\ra\la23\ra[\kf1l_2]}.
  \label{eq:C323Lint}
\end{align}
We are then left to feed in the parametrisation of \eqref{eq:lparam}, expand around $t=\infty$ and extract the coefficient of $t^0$. This analytic 
form still contains the dependence on $\gamma_\pm$ and we must still perform the sum before we arrive at the full scalar triangle coefficient. In 
principle this can be done numerically but in this case we find the analytic form after the $t$-expansion is rather lengthy and we can improve the 
situation considerably by performing the sum analytically. The procedure is straightforward but tends to lead us through relatively large intermediate 
expressions. We begin by removing all ``flatted" spinors in place of an explicit polynomial in $\gamma$. To do this, we make sure not to introduce 
spurious denominators by making use of:
\begin{align}
  \AB{2}{\kf1}{3} = -\frac{\gamma}{\gamma-m^2}\AB213.
\end{align}
The sum of the two solutions $\gamma_\pm$ is obtained from this polynomial form by rearrangement into a function of:
\begin{align}
  \gamma_+ + \gamma_- &= 2p_1\cdot p_4, & \gamma_+\gamma_- = m^4.
\end{align}
We are then left to partial fraction the spinor products. After the dust clears, we are left with a relatively compact form:
\begin{align}
  &iC^{[L]}_{3;4|1|23}(1_t^+,2^+,3^-,4_\tb^+) =
  \nonumber\\&
  \frac{6(\la\eta_{4}\eta_{1}\ra\la23\ra-2\la2\eta_{4}\ra\la3\eta_{1}\ra)\la3|1|2]^2m^3}{\beta^4s_{23}{}^2\la\eta_{1}\fl{1}\ra\la\eta_{4}\fl{4}\ra\la23\ra}
  \nonumber\\&
  +\frac{2(\la2\eta_{4}\ra\la3\eta_{1}\ra-\la\eta_{4}\eta_{1}\ra\la23\ra)\left(s_{23}+2\la2|1|2]\right)\la3|1|2]m^3}{\beta^2s_{23}\la\eta_{1}\fl{1}\ra\la\eta_{4}\fl{4}\ra\la23\ra\la2|1|3]}
  \nonumber\\&
  -\frac{2\left(2m^2+\la2|1|2]\right)\la2\eta_{1}\ra\la2\eta_{4}\ra\la3|1|2]m^3}{\beta^2\la\eta_{1}\fl{1}\ra\la\eta_{4}\fl{4}\ra\la23\ra\la2|1|3]^2}
%  \nonumber\\&
  +\frac{\left(s_{23}+2\la2|1|2]\right)\la2\eta_{1}\ra\la3\eta_{4}\ra[32]m^3}{\la\eta_{1}\fl{1}\ra\la\eta_{4}\fl{4}\ra\la2|1|3]^2}
  \nonumber\\&
  +\frac{\la2\eta_{1}\ra\la2\eta_{4}\ra\la3|1|2]^2m}{2\la\eta_{1}\fl{1}\ra\la\eta_{4}\fl{4}\ra\la23\ra\la2|1|3]}
%  \nonumber\\&
  -\frac{\la3\eta_{1}\ra\la3\eta_{4}\ra\la3|1|2]m}{2\la\eta_{1}\fl{1}\ra\la\eta_{4}\fl{4}\ra\la23\ra}
%  \nonumber\\&
  +\frac{s_{12}\la\eta_{4}\eta_{1}\ra\la3|1|2]m}{\la\eta_{1}\fl{1}\ra\la\eta_{4}\fl{4}\ra\la2|1|3]}
  \nonumber\\&
  -\frac{(\la3\eta_{1}\ra\la3\eta_{4}\ra\la2|1|3]-\la2\eta_{1}\ra\la2\eta_{4}\ra\la3|1|2])\la3|1|2]m}{4\beta^2\la\eta_{1}\fl{1}\ra\la\eta_{4}\fl{4}\ra\la23\ra\la2|1|3]}
  \nonumber\\&
  +\frac{3\left(8\la2|1|2]m^2+s_{23}^2\right)(\la3\eta_{1}\ra\la3\eta_{4}\ra\la2|1|3]-\la2\eta_{1}\ra\la2\eta_{4}\ra\la3|1|2])\la3|1|2]m}{4\beta^4s_{23}{}^2\la\eta_{1}\fl{1}\ra\la\eta_{4}\fl{4}\ra\la23\ra\la2|1|3]}
  \nonumber\\&
  +\frac{\la2\eta_{1}\ra\la2\eta_{4}\ra\la2|1|2]\la3|1|2][32]m}{2\la\eta_{1}\fl{1}\ra\la\eta_{4}\fl{4}\ra\la2|1|3]^2}
%  \nonumber\\&
  +\frac{\left(2m^2+\la3|1|3]\right)\la3\eta_{1}\ra\la3\eta_{4}\ra[32]m}{2\la\eta_{1}\fl{1}\ra\la\eta_{4}\fl{4}\ra\la2|1|3]}
  \nonumber\\&
  +\frac{\left(2m^2+\la2|1|2]\right)s_{23}\la2\eta_{1}\ra\la2\eta_{4}\ra\la2|1|2][32]m}{2\la\eta_{1}\fl{1}\ra\la\eta_{4}\fl{4}\ra\la2|1|3]^3}
%  \nonumber\\&
  -\frac{\beta^2s_{12}s_{23}^2\la\eta_{4}\eta_{1}\ra 
  m}{2\la\eta_{1}\fl{1}\ra\la\eta_{4}\fl{4}\ra\la2|1|3]^2},
\label{eq:C323L}
\end{align}
where $\beta=\sqrt{1-\tfrac{4m^2}{s_{23}}}$. Although one may not consider this form particularly elegant we note that it is much shorter than 
expression before the $\gamma$ sum. It is also among the most complicated coefficients that we encountered in the computation.

\subsubsection{Bubble Coefficients}

In this section we apply the Taylor expansion method to the computation of the bubble coefficient 
$C^{[L]}_{2;12|34}(1_t^+,2^+,3^+,4_\tb^+)$. The on-shell constraints can be solved in general leaving
two free complex parameters which we label $t$ and $y$. Owing to the two dimensional complex 
integration the pole structure is rather more involved than the triple cut considered above and we must consider 
triple cut contributions as well as the double cut in order compute the full coefficient \footnote{The alternative method of spinor 
integration produces an identical set of poles as can be seen from Mastrolia's evaluation of the double cut via Stoke's theorem 
\cite{Mastrolia:2009dr}.}. Our coefficient can therefore be written as:
\begin{align}
  &C_{2;12;34}(1_t^+,2^+,3^+,4_\tb^+)  = 
  -i\,{\rm inf}_t[{\rm inf}_y [\tilde{C}_{2;12|34}(1_t^+,2^+,3^+,4_\tb^+)]]\nonumber\\&
  -\frac{1}{2} \sum_{y_\pm} {\rm inf}_t [\tilde{C}_{3;12|3|4}(1_t^+,2^+,3^+,4_\tb^+)]]
  -\frac{1}{2} \sum_{y_\pm} {\rm inf}_t [\tilde{C}_{3;1|2|34}(1_t^+,2^+,3^+,4_\tb^+)]],
\end{align}
where we choose to parametrise the loop momentum as,
\begin{align}
  l_1^\mu = 
  y \kfm1 - \frac{\AB212-y s_{12}}{\AB212} p_3^\mu
  + \frac{t}{2} \AB{\kf1}{\gamma^\mu}{3}
  + \frac{y\left( \AB212-y s_{12} \right)}{2t\AB212} \AB{3}{\gamma^\mu}{\kf1}
\end{align}
with $\kf1=p_4-\frac{m^2}{\AB212}p_3$. In fact, with this choice we find the second term above 
vanishes. This is particularly convenient analytically, however, for a direct numerical evaluation, an 
independent massless vector would yield more stable results. Using the compact
expressions for the tree level amplitudes and summing over the internal helicities gives the 
integrands for the two non-zero contributions to be,
\begin{align}
  \tilde{C}_{2;12|34}&(1_t^+,2^+,3^+,4_\tb^+) = \nonumber\\&
 \frac{m^3 \left([l_12]^2 \la l_1|4|3]^2+[l_13]^2 \la l_1|1|2]^2\right) \la \eta_{1}\eta_{4}\ra }
 {\la 2|1|2]^2 \la 13\ra  [l_12] [l_13] \la 2 l_1\ra  \la \eta_{1}\fl{1}\ra  \la \eta_{4}\fl{4}\ra }
   +\frac{m^3 [23] \la l_1\eta_{1}\ra  \la l_1\eta_{4}\ra }{\la 2|1|2] \la
   l_12\ra  \la l_13\ra  \la \eta_{1}\fl{1}\ra  \la \eta_{4}\fl{4}\ra },
\end{align}
and,
\begin{align}
  &\tilde{C}_{3;1|2|34}(1_t^+,2^+,3^+,4_\tb^+) = 
\frac{i m [l_12]^3 \la l_1|4|3] \la 3|1l_3|3\ra  ([l_13] \la l_1\eta_{1}\ra  \la l_1\eta_{4}\ra +\la l_1|4|3] \la \eta_{1}\eta_{4}\ra )}
{\la 2|1|2] [l_13] \la l_13\ra  \la 3|l_3|l_1] \la \eta_{1}\fl{1}\ra  \la \eta_{4}\fl{4}\ra }\nonumber\\&
-\frac{i m^3 \la 3|1|2]^3 [l_13] \la l_12\ra ^2 \la \eta_{1}\eta_{4}\ra }
{\la 2|1|2] \la l_13\ra  \la 2|l_34|3\ra  \la 3|l_34|3\ra \la \eta_{1}\fl{1}\ra  \la 
\eta_{4}\fl{4}\ra }
+\frac{i m^3 [l_13] (\la \eta_{1}|l_3|2] \la \eta_{4}|l_3|2]-[2|l_31|2] \la \eta_{1}\eta_{4}\ra)}
{\la 2|1|2] [l_12] \la l_13\ra  \la 3|l_3|l_1] \la \eta_{1}\fl{1}\ra  \la \eta_{4}\fl{4}\ra }.
\end{align}
Just as in the three-mass triangle we find a considerable benefit from taking the extra effort to find a closed form for the triangle contributions which 
are free of square roots. Many terms cancel between the double and triple cuts so that the final result is simply:
\begin{align}
  -iC^{[L]}_{2;12|34}(1_t^+,2^+,3^+,4_\tb^+) =
  \frac{m^3[32](\la\eta_{1}|(1+2)(2+3)|\eta_{4}\ra)-s_{12}m^2}{\la2|1|2]^2\la\eta_{1}\fl{1}\ra\la\eta_{4}\fl{4}\ra\la23\ra}.
\end{align}

\subsubsection{Tadpole and On-Shell Bubble Coefficients \label{sec:tad}}

The computation of the tadpole coefficients directly from unitarity is complicated by the fact that 
the wave-function renormalisation contributions cause the double cuts to diverge. A way around this 
problem has been introduced in the context of a numerical application of $D$-dimensional generalised 
unitarity in reference \cite{Ellis:ttggg}. The method should apply equally well to analytic 
evaluation, but since it breaks gauge invariance, it can lead to large intermediate expressions. Such 
cuts also require the six-point tree-level amplitudes rendering the computation more difficult. 
Nevertheless, methods using spinor integration technique have been proposed 
\cite{Britto:2009wz,Britto:2010um}.

Both the tadpole and on-shell bubble coefficients combine to give the coefficient of the $\log(m^2)$ 
contribution to the full amplitude. The coefficient of this logarithm is completely fixed by the 
universal IR constraints once combined with the knowledge of the cuts considered in 
the previous section. With this in mind, we rearrange the integral basis as,
\begin{align}
  A^{[X]}_{4}&(1,2,3,4) = C^{[X]}_{4;1|2|3|4} \, I^{[X]}_{4;1|2|3|4} \nonumber\\&
  + \sum_{i=1}^{2}\sum_{j=i+1}^{3}\sum_{k=j+1}^{i-1}
  C^{[X]}_{3;i\dots j-1|j\dots k-1|k\dots i-1} \, I^{[X]}_{3;i\dots j-1|j\dots k-1|k\dots i-1}\nonumber\\&
  + \sum_{i=1}^{2}\sum_{j=i+2}^{i-2} 
  C^{[X]}_{2;i\dots j-1|j\dots i-1} \, I^{[X]}_{2;i\dots j-1|j\dots i-1} 
  + C^{[X]}_{2;m^2} I^{[X]}_{2;m^2} + R^{'[X]}+\mathcal{O}(\eps).
  \label{eq:scalar2}
\end{align}
For the gluon fusion channel the universal poles structure implies,
\begin{align}
  C^{[L]}_{2;m^2} &= - C^{[L]}_{2;12|34} - C^{[L]}_{2;23|41} + \frac{1}{2} A^{(0)}_4,\\
  C^{[R]}_{2;m^2} &= - C^{[R]}_{2;12|34} - C^{[R]}_{2;23|41} + \frac{1}{2} A^{(0)}_4,\\
  C^{[H]}_{2;m^2} &= - C^{[H]}_{2;23|41}
  \label{eq:wfcoeffsgg}
\end{align}
and the quark annihilation channel,
\begin{align}
  C^{[lc]}_{2;m^2} &= - C^{[lc]}_{2;12|34} - C^{[lc]}_{2;12|34} + \frac{8}{3} A^{(0)}_4,\\
  C^{[slc]}_{2;m^2} &= - C^{[slc,m]}_{2;23|41} - C^{[slc,0]}_{2;23|41} - A^{(0)}_4,\\
  C^{[H]}_{2;m^2} &= - C^{[H]}_{2;23|41} + \frac{2}{3} A^{(0)}_4.
  \label{eq:wfcoeffsqq}
\end{align}
In this case we returned to a Feynman based computation of the rational terms and remaining tadpole 
contributions. Owing to the simplified form of the reduction algorithm, and the fact that such 
contributions are independently gauge invariant, the approach gives a simple way to reach a compact  
form for the full amplitude. The final results for the tadpole coefficients turn out to be 
remarkably simple and suggestive that an approach based on matching with the universal IR and UV 
structure, as proposed in ref. \cite{Bern:massuni}, would generalise to all massive amplitudes. A 
complete description, however, remains for future study.

\subsection{Feynman Diagram based approach}

To compute rational contribution and provide numerical cross-checks for cut-constructible parts of the amplitude
we have performed two independent calculations based on the traditional Feynman diagram approach.

The diagrams were generated with DIANA \cite{Fleischer:2000zr} and then
further processed analytically with two independent FORM \cite{Vermaseren:2000nd} codes, to generate tensor
integral representations of the colour ordered one-loop amplitude.
The tensor integrals were reduced to scalar boxes, triangles and bubbles in $4-2\epsilon$ dimensions
using standard Passarino-Veltman reduction
in one case and to scalar integrals in shifted dimension according to
\cite{Davydychev:1991va,Fleischer:1999hq,Diakonidis:2009fx,Fleischer:2010sq} in the second case.
We performed calculations using both Dirac-spinor and spinor helicity methods in
Four-Dimensional-Helicity (FDH) and 't Hooft-Veltman (HV) schemes,
which allowed us to verify that scheme dependence is in agreement with predictions of \cite{Catani:1lmasssing}.

To maintain amplitude invariance under gauge transformations of gluon fields
it is necessary to perform finite renormalisation of the heavy quark mass \cite{Ellis:ttggg}.
This is done by including mass counter-term diagrams with the mass renormalisation defined by,
\begin{equation}
  m = (1+\delta Z_m) m^R,
  \label{eq:Zmdef}
\end{equation}
where
\begin{equation}
  \delta Z_m = -g_s^2 c_\Gamma C_F \left( \frac{3}{\eps} + 3\log\left(\frac{\mu_R^2}{m^2}\right) + 5 
  \right).
  \label{eq:Zmres}
\end{equation}

\section{Pole Structure \label{sec:IR}}

We have verified that our amplitudes satisfy the well known universal Infra-Red and Ultra-Violet 
pole structures \cite{Catani:1lmasssing}. These can be broken down into the 
contributions from each primitive amplitude \cite{Ellis:ttggg}. Representing the divergent parts of 
the amplitude by the function $V^{[X]}$ we can write,
\begin{equation}
	A_4^{[X]} = V^{[X]} A_4^{(0)} + F^{[X]}.
\end{equation}
For the gluon channel we find ($\beta$ as defined after eq. \eqref{eq:C323L}):
\begin{align}
	V_4^{[L]}(1_t,2,3,4_\tb) &= 
	- \frac{2}{\eps^2} 
	+ \frac{1}{2\eps} 
	- \frac{1}{\eps} \log\left( \frac{\mu_R^2 m^2}{\AB212^2} \right)
	- \frac{1}{\eps} \log\left( -\frac{\mu_R^2}{s_{23}} \right)
	\label{gg:poleL},\\
	V_4^{[R]}(1_t,2,3,4_\tb) &= 
	 \frac{1}{2\eps} 
	- \frac{1}{\eps}\frac{s_{23}-2m^2}{s_{23}\beta} \log\left( \frac{1-\beta}{1+\beta} \right)
	\label{gg:poleR}.
\end{align}
For the sub-leading colour we list the poles of the full colour ordered amplitude which is proportional to 
$A_4^{(0)}(1_t,2,3,4_\tb)+A_4^{(0)}(1_t,3,2,4_\tb) = A_4^{(0)}(1_t,2_\gamma,3_\gamma,4_\tb)$,
\begin{align}
	V^{[slc]}(1_t,2,3,4_\tb) = 
	&\frac{\AB212}{s_{23}\eps} \log\left( \frac{\mu_R^2 m^2}{\AB212^2} \right)
	+ \frac{\AB313}{s_{23}\eps} \log\left( \frac{\mu_R^2 m^2}{\AB313^2} \right)\nonumber\\
	+ &\frac{1}{\eps} \log\left( -\frac{\mu_R^2}{s_{23}} \right)
	+ \frac{1}{\eps}\frac{s_{23}-2m^2}{s_{23}\beta} \log\left( \frac{1-\beta}{1+\beta} \right)
	\label{gg:poleslc}.
\end{align}
The poles of full colour and helicity summed interference with the tree level can be written in terms 
of spin correlated Born amplitudes,
\begin{align}
  2\sum_{c,h} \mathcal{A}^{(1)}\cdot[\mathcal{A}^{(0)}]^* = &
	  2N_c\sum_{c,h} \bigg\{ |\mathcal{A}^{(0)}(2,3)|^2 \, V^{[L]}(1_t,2,3,4_\tb)
	+ |\mathcal{A}^{(0)}(3,2)|^2 \, V^{[L]}(1_t,3,2,4_\tb) \nonumber\\&
	- \frac{1}{N_c^2}|\mathcal{A}_4^{(0)}|^2 \, V^{[R]}
	- |\mathcal{A}_{4;\gamma}^{(0)}|^2 \, V^{[slc]} \bigg\},
\end{align}
where
\begin{align}
  \sum_{c,h}& |\mathcal{A}_4^{(0)}(2,3)|^2 = \nonumber\\&\frac{N_c^2-1}{N_c} \sum_{h} (N_c^2-1)|A_4^{(0)}(1_t,2,3,4_\tb)|^2 
					   + A_4^{(0)}(1_t,2,3,4_\tb)[A_4^{(0)}(1_t,3,2,4_\tb)]^*,
\end{align}
and $\sum_{c,h} |\mathcal{A}_4^{(0)}(2,3)|^2+|\mathcal{A}_4^{(0)}(3,2)|^2  = \sum_{c,h} |\mathcal{A}_4^{(0)}|^2$. Turning our attention to the quark channel the 
analogous structure can be written as:
\begin{align}
	V^{[lc]}(1_t,2_\qb,3_q,4_\tb) &= 
	\frac{1}{\eps^2} 
	+ \frac{8}{3\eps} 
	- \frac{1}{\eps} \log\left( \frac{\mu_R^2 m^2}{\AB212^2} \right)
	\label{qq:polelc},\\
	V^{[slc]}(1_t,2_\qb,3_q,4_\tb) &= 
	\frac{1}{\eps^2} 
	- \frac{1}{\eps} 
	- \frac{1}{\eps} \log\left( -\frac{\mu_R^2}{s_{23}} \right)
	- \frac{1}{\eps}\frac{s_{23}-2m^2}{s_{23}\beta} \log\left( \frac{1-\beta}{1+\beta} \right)
	\label{qq:poleslc},\\
	V^{[f]}(1_t,2_\qb,3_q,4_\tb) &= 
	V^{[f]}(1_t,2_\qb,3_q,4_\tb) = \frac{2}{3\eps} 
	\label{qq:polef},
\end{align}
\begin{align}
	2\sum_{c,h} \mathcal{A}_4^{(1)}\cdot&[\mathcal{A}_4^{(0)}]^* = 
	  2\sum_{c,h} |\mathcal{A}_4^{(0)}|^2 \bigg\{  
	  N_c V^{[lc]}(1_t,2_\qb,3_q,4_\tb) - (N_f+N_H) V^{[f]} \nonumber\\& 
	  - \frac{2}{N_c}\left( V^{[lc]}(1_t,2_\qb,3_q,4_\tb) - V^{[lc]}(1_t,3_\qb,2_q,4_\tb) \right)
	  - \frac{1}{N_c} V^{[slc]}(1_t,2_\qb,3_q,4_\tb)  \bigg\}.
\end{align}

\section{Renormalisation and Scheme Dependence \label{sec:renorm}}

The one-loop amplitudes presented in the previous sections still contain UV divergences which need 
to be renormalised.  At this point we want to remind the reader that we already included the mass 
renormalisation for the top quark mass, which we define using the pole scheme. Thus we are left with the 
wave function renormalisation and the running of the strong coupling constant. For the former we  
are using an on-shell prescription which takes into account all self energy contributions, whereas the coupling 
constant is renormalised in the $\overline{\rm MS}$ scheme. To decouple the top quark from the running of $\alpha_s$,
 one subtracts the diagrams with a top quark in the loop at zero 
momentum transfer. Following the \cite{Beenakker:2002nc,Harris:2002md} we define the renormalisation 
constants for the gluon and fermion fields and that of the strong coupling,
\begin{align}
  G_{a,\mu}^{0} &= (1 + \delta Z_G/2) G_{a,\mu}\,, &
  \psi_Q^0 &= (1 + \delta Z_Q/2) \psi_Q\,, &
  (g_s^0)^2 = (1 + \delta Z_{g_s}) g_s^2\,,
\end{align}
where, in the FDH scheme,
\begin{align}
  \delta Z_Q &= \delta Z_m\, , \qquad \delta Z_G = -\frac{2g_s^2 c_\Gamma}{3 \eps} \left( 
  \frac{\mu_R^2}{m^2} 
 \right)^\eps\, , \\
 \delta Z_{g_s} &=  -g_s^2 c_\Gamma \left\{\frac{1}{\eps} \left( \frac{11}{3} N_c - \frac{2}{3} N_f - \frac{2}{3} \left( 
 \frac{\mu_R^2}{m^2} \right)^\eps \right) - \frac{N_c}{3} \right\} \, ,
\end{align}
with $C_F = \frac{N_c^2 -1}{2 N_c}$. Details of the scheme dependence of the coupling constant renormalisation 
can be found in reference \cite{Kunszt:1l2to2}. Summing up all contributions the renormalised amplitudes are given by,
\begin{align}
  &\mathcal{A}_4^{(1), {\rm ren}} \left(1_t, 2, 3, 4_\tb \right) =
  \mathcal{A}_4^{(1)} \left(1_t,2,3,4_\tb \right) +
  \left(\delta Z_G + \delta Z_Q + \delta Z_{g_s}  \right) \mathcal{A}_4^{(0)} \left(1_t,2,3,4_\tb \right) \nonumber\\&
= \mathcal{A}_4^{(1)} \left(1_t,2,3,4_\tb \right) - g_s^2 c_\Gamma\mathcal{A}_4^{(0)} \left(1_t,2,3,4_\tb \right)\times\nonumber\\& \bigg\{ 
\frac{11N_c-2N_f}{3\eps} - \frac{N_c}{3} + C_F\left( \frac{3}{\eps} + 3\log\left( 
\frac{\mu_R^2}{m^2} \right) + 5 \right)\bigg\}, \\
&\mathcal{A}_4^{(1), {\rm ren}} \left(1_t, 2_\qb, 3_q, 4_\tb \right) = \mathcal{A}_4^{(1)} 
\left(1_t,2_\qb,3_q,4_\tb \right) 
+ \left( \delta Z_Q + \delta Z_{g_s}\right)\mathcal{A}_4^{(0)} \left(1_t,2_\qb,3_q,4_\tb \right) \nonumber\\&
= \mathcal{A}_4^{(1)} \left(1_t,2_\qb,3_q,4_\tb \right) - g_s^2 c_\Gamma \mathcal{A}_4^{(0)} \left(1_t,2_\qb,3_q,4_\tb \right) 
\times\nonumber\\&
\bigg\{
\frac{11N_c-2N_f}{3\eps} - \frac{N_c}{3}
+ C_F\left( \frac{3}{\eps} + 3\log\left( \frac{\mu_R^2}{m^2} \right) + 5 \right)
- \frac{2}{3\eps} - \frac{2}{3}\log\left( \frac{\mu_R^2}{m^2} \right)
\bigg\}.
\end{align}
To convert these renormalised amplitudes to the 't Hooft-Veltman (HV) scheme we follow the well known 
universal structure \cite{Kunszt:1l2to2,Catani:1lmasssing,Beenakker:2002nc,Harris:2002md}. For our 
amplitudes this can be summarised as,
\begin{align}
\mathcal{A}_4^{(1), \, [\text{HV}] } \left(1_t,2,3,4_\tb \right) &=
\mathcal{A}_4^{(1)} \left(1_t,2,3,4_\tb \right) - g_s^2 c_\Gamma C_F \mathcal{A}_4^{(0)} \left(1_t,2,3,4_\tb \right), \\  
\mathcal{A}_4^{(1), \, [\text{HV}] } \left(1_t,2_\qb,3_q,4_\tb \right) &=
\mathcal{A}_4^{(1)} \left(1_t,2_\qb,3_q,4_\tb \right) - g_s^2 c_\Gamma \left( 2 C_F - \frac{N_c}{3} \right) 
\mathcal{A}_4^{(0)} \left(1_t,2_\qb,3_q,4_\tb  \right).
\end{align}
After renormalisation, it becomes,
\begin{align}
\mathcal{A}_4^{(1), \, [\text{HV}], {\rm ren} } \left(1_t,2,3,4_\tb \right) &=
\mathcal{A}_4^{(1), {\rm ren}} \left(1_t,2,3,4_\tb \right) - g_s^2 c_\Gamma \frac{N_c}{3} \mathcal{A}_4^{(0)} \left(1_t,2,3,4_\tb \right), \\  
\mathcal{A}_4^{(1), \, [\text{HV}], {\rm ren} } \left(1_t,2_\qb,3_q,4_\tb \right) &=
\mathcal{A}_4^{(1), {\rm ren}} \left(1_t,2_\qb,3_q,4_\tb \right) - g_s^2 c_\Gamma C_F
\mathcal{A}_4^{(0)} \left(1_t,2_\qb,3_q,4_\tb  \right).
\end{align}

\section{Tree-level amplitudes \label{sec:treeamps}}
 
The tree-level amplitudes can be computed using BCFW recursion relations 
\cite{Britto:rec,Britto:recproof} together with the spinor conventions of the previous section.

The three-point amplitudes are calculated directly from the Feynman vertex using arbitrary reference vectors.
As an example we consider the 
\begin{align}
	A_3(1_t^{+},2^{\lambda_2},3_\tb^{+}) 
	= \frac{i}{\sqrt{2}}\,\overline{u_{+}}(1,m;\fl1,\eta_1) \slashed{\varepsilon}_{\lambda_2}(2,\xi) v_{+}(3,m;\fl3,\eta_3),
	\label{eq:3ptdef}
\end{align}
which can be expanded to give
\begin{align}
  A_3(1_t^+,2^{\lambda_2},3_\tb^+) &= \frac{i}{\sqrt{2}}\, \frac{ \la \eta_1 | (\slashed{1}+m)
  \slashed{\varepsilon}_{\lambda_2}(2,\xi) (\slashed{3}-m)| \eta_3 \ra}{\A{\eta_1}{\fl 1}\A{\fl
	3}{\eta_3}}.
\end{align}
The tree-level helicity amplitudes, expressed in terms of spinor products, are listed in 
Appendix~\ref{App:trees}.

\section{One-loop Amplitudes \label{sec:1lamps}}

In this section we present a complete set of one-loop helicity amplitudes needed for $t\tb$ 
production at hadron colliders.

\subsection{Notation and Integral Functions}

We define the general scalar integral as \footnote{This follows the conventions of {\tt qcdloop} which was used for numerical evaluations 
\cite{Ellis:scalarints}.}
\begin{equation}
  I_n = \frac{\mu_R^{2\eps}\Gamma(1-2\eps)}{i(\pi)^{2-\eps}\Gamma(1-\eps)^2\Gamma(1+\eps)} \int d^Dl 
  \frac{1}{\prod_{i=1}^n (l-k_i)^2-m_i^2},
\end{equation}
$\{k_i\}$ are the sums of external momenta entering each of the $n$ vertices of the graph.
We will denote the basis integrals for our helicity amplitudes as,
\begin{align}
  I_{4;1|2|3|4} &= I_4(m^2,0,0,m^2,s_{12},s_{23},m^2,0,0,0) \\
  I^m_{4;1|2|3|4} &= I_4(m^2,0,0,m^2,s_{12},s_{23},0,m^2,m^2,m^2) \\
  I_{4;1|2|4|3} &= I_4(m^2,0,m^2,0,s_{12},s_{13},0,m^2,m^2,0) \\
  I_{3;12|3|4} &= I_3(s_{12},0,m^2,m^2,0,0) \\
  I^m_{3;12|3|4} &= I_3(s_{12},0,m^2,0,m^2,m^2) \\
  I_{3;13|2|4} &= I_3(s_{13},0,m^2,m^2,0,0) \\
  I^m_{3;13|2|4} &= I_3(s_{13},0,m^2,0,m^2,m^2) \\
  I_{3;2|3|41} &= I_3(s_{23},0,0,0,0,0) \\
  I^m_{3;2|3|41} &= I_3(s_{23},0,0,m^2,m^2,m^2) \\
  I_{3;1|23|4} &= I_3(s_{23},m^2,m^2,0,0,m^2) \\
  I^m_{3;1|23|4} &= I_3(s_{23},m^2,m^2,m^2,m^2,0).
  \label{eq:intnotdef}
\end{align}
As described in section \ref{sec:tad} we find it convenient to move all $\log(m^2)$ dependence into the on-shell bubble. Therefore our helicity 
amplitudes are written in terms of the following two-point functions,
\begin{align}
  F_{2;12} &= I_2(s_{12},0,m^2) - I_2(m^2,0,m^2) \,, \\
  F^m_{2;23} &= I_2(s_{23},m^2,m^2) - I_2(0,m^2,m^2) \,, \\ 
  \hat{I}_{2;23} &= I_2(s_{23},0,0) - I_2(m^2,0,m^2) + 2 +\mathcal{O}(\eps) \,, \\
  I_{2;m} &= I_2(m^2,0,m^2) \,, \\
  I^m_{2;m} &= I_2(0,m^2,m^2) \,,
\end{align}
and the tadpole function is removed via
\begin{equation}
  I_1(m^2) = m^2 \, \left( I_2(m^2,0,m^2) - 1 \right) + \mathcal{O}(\eps)\, .  
\end{equation}
Finite box functions are used to make the IR poles explicit:
\begin{equation}
  F^m_{4;1|2|3|4} = I^m_{4;1|2|3|4} - \frac{1}{\AB{2}{1}{2}} I^m_{3;1|23|4}.
\end{equation}
We also find benefits in cancellation of spurious poles that can be made explicit through the use of 
higher dimensional integral functions:
\begin{align}
  &I_{4;1|2|3|4}^{6-2\eps} = \frac{1}{2\AB213\AB312}\Big(  
  {-}\AB212^2s_{23} I_{4;1|2|3|4}\nonumber\\&
  +\left( \AB212+2m^2 \right)s_{23} I_{3;1|23|4}
  +2\AB212^2 I_{3;12|3|4}
  +\AB212s_{23} I_{3;2|3|41}
  \Big),
  \\
  &I_{4;1|2|3|4}^{m,6-2\eps} = \frac{1}{2\AB213\AB312}\Big(  
  {-}\AB212^2s_{23}\beta^2 I^m_{4;1|2|3|4}\nonumber\\&
  +\AB212s_{23}\beta^2 I^m_{3;1|23|4}
  +2\AB212\left( \AB212+2m^2 \right) I^m_{3;12|3|4}
  +s_{23}\left( \AB212+2m^2 \right) I^m_{3;2|3|41}
  \Big).
  \label{eq:6dboxes}
\end{align}
\def\ampnorm{\A{\eta_1}{\fl1}\A{\eta_4}{\fl4}}
\def\nnA{\tilde{A}}
Since all the expressions are quoted for the same configuration of heavy quark helicities we find it 
convenient to factor out an overall normalisation $\ampnorm$ and define,
\begin{equation}
  \nnA_4^{(0)} = -i\ampnorm A_4^{(0)}.
\end{equation}

\subsection{Primitive amplitudes for $gg\to t\tb$}
\vspace{-0.7cm}
\begin{align}
 \begin{split}
-i&\ampnorm A_4^{[L]} \left(1_t^{+},2^{+},3^{+},4_\tb^{+} \right)= \\
& I_{4;1|2|3|4} \left( m^3 \A{\eta_1}{\eta_4} \B{2}{3}^2 \right)
 - F_{2;12} \bigg( \frac{m^3 \B{2}{3} (2 \A{\eta_1}{\eta_4} s_{12} - \AA{\eta_1}{K_{12} K_{23}}{\eta_4})}{\A{2}{3} \AB{2}{1}{2}^2} \bigg)\\
& + \frac{1}{2} \nnA_4^{(0)} \left(1_t^{+},2^{+},3^{+},4_\tb^{+} \right) \times \left( I_{2;m} - 1 \right)  \\
& -  \bigg( \frac{m \B{2}{3} (\A{\eta_1}{\eta_4} \AB{2}{1}{2} + \AA{\eta_1}{K_{12} K_{23}}{\eta_4})}{2 \A{2}{3} \AB{2}{1}{2}} -\frac{m (\A{\eta_1}{\eta_4} \AB{2}{1}{2} - \A{2}{\eta_1} \A{3}{\eta_4} \B{2}{3})}{3 \A{2}{3}^2} \bigg) 
 \end{split}
\end{align}

\begin{align}
 \begin{split}
   i&\ampnorm A_4^{[R]} \left(1_t^{+},2^{+},3^{+},4_\tb^{+} \right)=\\
& - F_{4;1|2|3|4}^{m} \bigg(
     \frac{\A{3}{\eta_1} \A{3}{\eta_4} \B{2}{3}^2 m^3 \left(2 m^2 + \AB{2}{1}{2}\right)}{2 \A{2}{3} \AB{3}{1}{2}} 
    - \frac{m^3 \A{2}{\eta_1} \A{2}{\eta_4} \B{2}{3}^2 \left(2 m^2 + \AB{2}{1}{2}\right)}{2 \A{2}{3} \AB{2}{1}{3}} \\ &
    + \frac{ m^3 \B{2}{3} \left( \A{\eta_1}{\eta_4} \left(2 m^2 - s_{23} \right) + \AA{\eta_1}{K_{12} K_{23} }{\eta_4} \right) }{\A{2}{3}} \bigg) \\
& - I_{3;1|23|4}^{m} \bigg(
	\frac{\left(2 m^2 - s_{23} \right) \A{\eta_1}{\eta_4} \B{2}{3} m^3}{\A{2}{3} \AB{2}{1}{2}} \bigg) \\
& + I_{3;12|3|4}^{m} \bigg(
      \frac{m^3 \B{2}{3} (2 \A{\eta_1}{\eta_4} \A{2}{3} + 4 \A{2}{\eta_4} \A{3}{\eta_1} )}{\A{2}{3}^2} 
      + \frac{m^3 \B{2}{3} \A{2}{\eta_1} \A{2}{\eta_4} \AB{2}{1}{2}}{\A{2}{3}^2 \AB{2}{1}{3}} \\ &
      + \frac{m^3 \B{2}{3} \left( \A{3}{\eta_1} \A{3}{\eta_4} \AB{2}{1}{3} - \A{2}{\eta_1} \A{2}{\eta_4} \AB{3}{1}{2} \right)}{\A{2}{3}^2 \AB{2}{1}{2}} 
      - \frac{m^3 \B{2}{3} \A{3}{\eta_1} \A{3}{\eta_4} \AB{2}{1}{2}}{\A{2}{3}^2 \AB{3}{1}{2}}
	\bigg) \\
& - I_{3;2|3|41}^{m} \bigg(
	\frac{m^3 \A{2}{\eta_1} \A{2}{\eta_4} \B{2}{3}^2 }{2 \A{2}{3} \AB{2}{1}{3}}
	- \frac{m^3 \A{3}{\eta_1} \A{3}{\eta_4} \B{2}{3}^2 }{2 \A{2}{3} \AB{3}{1}{2}}
	+ \frac{m^3 \A{\eta_1}{\eta_4} \B{2}{3} }{\A{2}{3}}
	\bigg) \\
& + F_{2;12} \bigg(
      \frac{m^3 \B{2}{3} \left(2 s_{12} \A{\eta_1}{\eta_4} - \AA{\eta_1}{K_{12} K_{23} }{\eta_4}\right)}{\A{2}{3} \AB{2}{1}{2}^2} \bigg) \\
& - \frac{1}{2} \nnA_4^{(0)}\left(1_t^{+},2^{+},3^{+},4_\tb^{+} \right) \left( I_{2;m} - 1 \right) 
 - \bigg( \frac{m \B{2}{3} \left(\AA{\eta_1}{K_{12} K_{23} }{\eta_4} +\A{\eta_1}{\eta_4} \AB{2}{1}{2} \right)}{2 \A{2}{3} \AB{2}{1}{2} } \bigg)
\\[4pt] & \text{where } K_{ij} = p_i+p_j.
 \end{split}
\end{align}

% where $K_{ij} = p_i+p_j$.
\vspace{-1cm}
\begin{align}
  \begin{split}
-i&\ampnorm A_4^{[L]} \left(1_t^{+},2^{+},3^{-},4_\tb^{+} \right)= \\
&  \nnA_4^{(0)}\left(1_t^{+},2^{+},3^{-},4_\tb^{+} \right) \times 
       \bigg(- s_{23} \AB{2}{1}{2} I_{4;1|2|3|4} + \frac{1}{2} I_{2;m} - \frac{\AB{2}{1}{2} + 2 m^2}{\AB{2}{1}{2}} F_{2;12} \bigg) \\  
& + \left( I_{3;1|23|4} + \frac{2}{s_{23}} \hat{I}_{2;23} - \frac{4}{s_{23}} \right) \times 
       \bigg( \frac{3 m \AB{3}{1}{2} \left( \A{3}{\eta_1} \A{3}{\eta_4} \AB{2}{1}{3} - \A{2}{\eta_1} \A{2}{\eta_4} \AB{3}{1}{2} \right)}{4 \A{2}{3} \AB{2}{1}{3} \beta^4} \bigg) \\ 
& + \left( I_{3;1|23|4} - \frac{4}{s_{23}} \right) \times 
       \bigg(
       \frac{6 m^3 \AB{2}{1}{2} \AB{3}{1}{2} \left(\A{3}{\eta_1} \A{3}{\eta_4} \AB{2}{1}{3} - \A{2}{\eta_1} \A{2}{\eta_4} \AB{3}{1}{2} \right)}{\A{2}{3} \AB{2}{1}{3} s_{23}^2 \beta^4} \\ &
      - \frac{6 m^3 \AB{3}{1}{2}^2 \left(\A{2}{\eta_4} \A{3}{\eta_1} + \A{2}{\eta_1} \A{3}{\eta_4}\right)}{\A{2}{3} s_{23}^2 \beta^4} \bigg) 
  - \frac{m \A{3}{\eta_1} \A{3}{\eta_4} \AB{3}{1}{2} \B{2}{3}}{2} I_{4;1|2|3|4} \\
&  +  \frac{m \A{3}{\eta_1} \A{3}{\eta_4} \AB{3}{1}{2} \B{2}{3}}{2 \AB{2}{1}{2} } I_{3;2|3|41} 
   - \frac{m \A{3}{\eta_1} \A{3}{\eta_4} \AB{3}{1}{2} }{ \A{2}{3}} I_{3;12|3|4} 
   + \left( \hat{I}_{2;23} - 2 \right) \\ & 
       \times \frac{m \A{2}{\eta_1} \A{3}{\eta_4} \AB{3}{1}{2} \left(2 \AB{2}{1}{2} + s_{23} \right) }{\A{2}{3} \AB{2}{1}{3} s_{23} \beta^2} 
+ I^{6-2\eps}_{4;1|2|3|4} \bigg(
    \frac{m^3 \A{3}{\eta_1} \A{3}{\eta_4} \AB{3}{1}{2} \B{2}{3}}{\AB{2}{1}{2}^2} \\ &
    + \frac{m \A{2}{\eta_1} \A{2}{\eta_4} \AB{3}{1}{2} \left(\AB{2}{1}{2} s_{23} + \AB{2}{1}{3} \AB{3}{1}{2} \right) }{\A{2}{3} \AB{2}{1}{3}^2}
    - \frac{2 m^3 \A{\eta_1}{\eta_4} \AB{3}{1}{2} }{\AB{2}{1}{3}} \\ &
    + \frac{m \A{\eta_1}{\eta_4} \AB{3}{1}{2} \left( \AB{2}{1}{3} \AB{3}{1}{2} - \AB{2}{1}{2}^2 \right)}{\AB{2}{1}{2} \AB{2}{1}{3} }
    + \frac{2 m^3 \A{2}{\eta_1} \A{3}{\eta_4} \AB{3}{1}{2} \B{2}{3}}{\AB{2}{1}{2} \AB{2}{1}{3} } 
    \bigg) \\ 
& + I_{3;1|23|4} \bigg(
\frac{m \AB{3}{1}{2}^2 \left(\A{\eta_1}{\eta_4} \A{2}{3} + 2 \A{2}{\eta_1} \A{3}{\eta_4} \right)}{\AB{2}{1}{2} \A{2}{3}} 
			+ \frac{m \A{2}{\eta_1} \A{3}{\eta_4} \AB{3}{1}{2} (\AB{2}{1}{2} + 2 m^2)}{\A{2}{3} \AB{2}{1}{3} \beta^2} \\ &
			- \frac{m \A{3}{\eta_1} \A{3}{\eta_4} \AB{3}{1}{2} \left( \AB{2}{1}{2}^2 - \AB{2}{1}{2} s_{23} + 2 m^2 s_{23} \right)}{2 \A{2}{3} \AB{2}{1}{2}^2 } 
     - \frac{m^3 \A{\eta_1}{\eta_4} \AB{3}{1}{2}}{\AB{2}{1}{3}} \\ &
     + \frac{m \A{2}{\eta_1} \A{3}{\eta_4} \AB{3}{1}{2} \left( \AB{2}{1}{2} + 2 s_{23} \right)}{\A{2}{3} \AB{2}{1}{3}}
     + \frac{m^3 \A{2}{\eta_1} \A{2}{\eta_4} \AB{3}{1}{2} \B{2}{3} (2 \AB{2}{1}{2} + s_{23})}{ \AB{2}{1}{3}^2 s_{23} \beta^2} \\ &
     + \frac{m \A{2}{\eta_1} \A{2}{\eta_4} \AB{3}{1}{2}^2}{2 \A{2}{3} \AB{2}{1}{3}}
     - \frac{m \AB{3}{1}{2} \B{2}{3} \left(\A{2}{\eta_1} \A{2}{\eta_4} \AB{3}{1}{2} - \A{3}{\eta_1} \A{3}{\eta_4} \AB{2}{1}{3} \right) }{4 \AB{2}{1}{3} s_{23} \beta^2}
       \bigg) \\
& + \hat{I}_{2;23} \bigg(
     \frac{m \AB{3}{1}{2} \left(\A{\eta_1}{\eta_4} \AB{2}{1}{2} - \A{2}{\eta_1} \A{3}{\eta_4} \B{2}{3}\right)}{\AB{2}{1}{3} s_{23}}
		 -  \frac{m \A{2}{\eta_1} \A{2}{\eta_4} \AB{3}{1}{2} \left(\AB{2}{1}{2} + 2 m^2 \right)}{\A{2}{3} \AB{2}{1}{3}^2 \beta^2} \\ &
     - \frac{12 m^2 + s_{23} \beta^2}{s_{23} \beta^2} \times \frac{m \AB{3}{1}{2}^2 \left( \A{2}{\eta_1} \A{3}{\eta_4} + \A{3}{\eta_1} \A{2}{\eta_4} \right) }{\A{2}{3} s_{23}^2 \beta^2}  \\ &  
     - \frac{12 m^2 + s_{23} \beta^2}{s_{23} \beta^2} \times \frac{m \AB{2}{1}{2} \AB{3}{1}{2} \left(\A{2}{\eta_1} \A{2}{\eta_4} \AB{3}{1}{2} - \A{3}{\eta_1} \A{3}{\eta_4} \AB{2}{1}{3} \right)}{\A{2}{3} \AB{2}{1}{3} s_{23}^2 \beta^2} \bigg) \\
& - F_{2;12} \bigg(
     \frac{m \A{\eta_1}{\eta_4} \AB{3}{1}{2} \left(\AB{2}{1}{2} + 2 m^2 \right)}{\AB{2}{1}{3} s_{23}} 
     - \frac{m \A{2}{\eta_1} \A{2}{\eta_4} \AB{2}{1}{2} \AB{3}{1}{2}}{\A{2}{3} \AB{2}{1}{3}^2} \\ &
     - \frac{m \A{3}{\eta_4} \AB{\eta_1}{1}{2} \AB{3}{1}{2}}{2 \AB{2}{1}{2} s_{12}}
     + \frac{m \A{3}{\eta_1} \A{3}{\eta_4} \AB{3}{1}{2} \left( 2 m^2 - \AB{2}{1}{2} \right)}{2 \A{2}{3} \AB{2}{1}{2}^2}
     + \frac{2 m^3 \A{2}{\eta_1} \A{3}{\eta_4} \AB{3}{1}{2}}{\A{2}{3} \AB{2}{1}{2} \AB{2}{1}{3}} \\ &
     + \frac{m \A{2}{\eta_1} \A{3}{\eta_4} \AB{3}{1}{2}^2}{\A{2}{3} \AB{2}{1}{2}^2} 
     - \frac{m \A{2}{\eta_1} \A{2}{\eta_4} \AB{3}{1}{2}^2 \left( 2 \AB{2}{1}{2}^2 + 2 \AB{2}{1}{3} \AB{3}{1}{2} + \AB{2}{1}{2} s_{23} \right)}{2 \A{2}{3} \AB{2}{1}{2}^2 \AB{2}{1}{3} s_{23}}
 \bigg) \\
& + \frac{m \A{3}{\eta_4} \AB{3}{1}{2} \left(\A{2}{\eta_1} \AB{3}{1}{2} - \A{3}{\eta_1} \AB{2}{1}{2} \right)}{2 \A{2}{3} \AB{2}{1}{2} s_{12}}
    - \frac{2 m \A{\eta_1}{\eta_4} \AB{2}{1}{2} \AB{3}{1}{2}}{\AB{2}{1}{3} s_{23}}
     - \frac{m \A{\eta_1}{\eta_4} \AB{3}{1}{2}^2}{2 \AB{2}{1}{2} s_{23}}
 \\ &
    + \frac{m \A{2}{\eta_1} \AB{3}{1}{2} \left(\A{2}{\eta_4} \AB{3}{1}{2} - 4 \A{3}{\eta_4} \AB{2}{1}{2} \right)}{2 \A{2}{3} \AB{2}{1}{2} \AB{2}{1}{3}} 
     - \frac{m \A{2}{\eta_1} \A{2}{\eta_4} \AB{3}{1}{2}^2 }{\A{2}{3} \AB{2}{1}{3} s_{23} \beta^2}
     + \frac{m \A{3}{\eta_1} \A{3}{\eta_4} \AB{3}{1}{2}}{\A{2}{3} s_{23} \beta^2}
%   \end{split}
%   \nonumber
%\end{align}
\\ &
% \begin{align}
%   \begin{split}
     + \frac{2 m \A{2}{\eta_1} \A{2}{\eta_4} \AB{3}{1}{2} (\AB{2}{1}{2} + 2 m^2)}{\A{2}{3} \AB{2}{1}{3}^2 \beta^2} 
\end{split}
\end{align}

\vspace{-1cm}
\begin{align}
 \begin{split}
i&\ampnorm A_4^{[R]} \left(1_t^{+},2^{+},3^{-},4_\tb^{+} \right) = \\
& \nnA_4^{(0)}\left(1_t^{+},2^{+},3^{-},4_\tb^{+} \right) \times \bigg(
        (s_{23}{-} 2 m^2) I_{3;1|23|4}^{m} 
	- \frac{2 m^2 \left(2 s_{12} - s_{23} \right)}{\AB{2}{1}{2} \beta^2} I_{3;2|3|41}^{m}
 	+ \frac{2 s_{12}}{\AB{2}{1}{2}} F_{2;12}
 	- \frac{1}{2} I_{2;m} 
	\bigg) %\\ &
     \\ &
 + \bigg( 
      {-}(\AB{2}{1}{2} + 2 m^2) \B{2}{3} I_{4;1|2|3|4}^{m,6-2\eps} 
     + \frac{2 m^2 \AB{2}{1}{2}}{\A{2}{3}} I_{3;12|3|4}^{m}  %\\ &
    - m^2 \B{2}{3} I_{3;2|3|41}^{m} + \frac{\AB{2}{1}{2} + 2 m^2}{\A{2}{3}} F_{2;12} \\ & 
		- \frac{\AB{2}{1}{2}}{\A{2}{3}} (F_{2;23}^{m} - 2) \bigg)  \times 
      \frac{m \A{2}{\eta_1} \AB{3}{1}{2} \left( \A{2}{\eta_4} \AB{2}{1}{2} + 2 \A{3}{\eta_4} \AB{2}{1}{3}\right)}{\AB{2}{1}{2} \AB{2}{1}{3}^2 }  
%
% \\ &
%
- I^{m,6-2\eps}_{4;1|2|3|4} \times \\ & \times\bigg(
	\frac{m^3 \A{2}{\eta_4} \A{3}{\eta_1} \AB{3}{1}{2}^2 }{\A{2}{3} \AB{2}{1}{2}^2 \beta^2} %\\ &
	+ \frac{m \AB{3}{1}{2} \B{2}{3} \A{3}{\eta_1} \A{3}{\eta_4} (\AB{2}{1}{2} + 2 m^2)}{\AB{2}{1}{2}^2} 
	+ \frac{m^3 \A{3}{\eta_1} \A{3}{\eta_4} \AB{3}{1}{2} \B{2}{3}}{2 \AB{2}{1}{2}^2} \\ &
	+ \frac{m \A{2}{\eta_1} \A{3}{\eta_4} \AB{3}{1}{2} \left(2 \AB{2}{1}{2}^2 + s_{23} (\AB{2}{1}{2} + 2 m^2) \right)}{\A{2}{3} \AB{2}{1}{2} \AB{2}{1}{3}} 
	+ \frac{m \A{2}{\eta_1} \A{2}{\eta_4} \AB{3}{1}{2}^2 \left(\AB{2}{1}{2} + 2 m^2\right)}{\A{2}{3} \AB{2}{1}{3} s_{23} \beta^2} \\ &
	+ \frac{m \A{2}{\eta_1} \AB{3}{1}{2}^2 \left(\A{2}{\eta_4} \AB{3}{1}{2} + \A{3}{\eta_4} \AB{2}{1}{2} \right)}{\A{2}{3} \AB{2}{1}{2} s_{23} \beta^2}
+ \frac{3 m \A{2}{\eta_1} \A{3}{\eta_4} \AB{3}{1}{2}^2  \left(\AB{2}{1}{2}^2 + s_{12} s_{23} \right)}{\A{2}{3} \AB{2}{1}{2}^2 s_{23} \beta^2} \\ &
	+ \frac{2 m \A{3}{\eta_1} \A{3}{\eta_4} \AB{3}{1}{2} \left(\AB{2}{1}{2} + 2 m^2\right)}{\A{2}{3} \AB{2}{1}{2} \beta^2} 
	+ \frac{m^3 \A{3}{\eta_1} \A{3}{\eta_4} \AB{3}{1}{2} \left(2 \AB{2}{1}{2} + s_{23}\right)}{2 \A{2}{3} \AB{2}{1}{2}^2 \beta^2} \\ &
	- \frac{m \A{\eta_1}{\eta_4} \AB{3}{1}{2} s_{12} \left(\AB{2}{1}{2}^2 - s_{12} s_{23} \right)}{\AB{2}{1}{3} \AB{2}{1}{2}^2}  
	\bigg) 
 + F_{4;1|2|3|4}^{m} \frac{m \A{3}{\eta_4} \AB{3}{1}{2} \left(2 \A{2}{\eta_1} \AB{3}{1}{2} + 3 m^2 \A{3}{\eta_1} \right)}{\A{2}{3}} \\
& + I_{3;12|3|4}^{m} \bigg( 
	\frac{5 m^3 \A{3}{\eta_1} \A{3}{\eta_4} \AB{3}{1}{2} }{\A{2}{3} \AB{2}{1}{2} } 
	- \frac{2 m^3 \A{\eta_1}{\eta_4} \AB{3}{1}{2} s_{12}}{\AB{2}{1}{2} \AB{2}{1}{3}} 
	- \frac{m^3 \A{3}{\eta_1} \A{3}{\eta_4} \AB{3}{1}{2} \left(8 \AB{2}{1}{2} + s_{23} \right)}{\A{2}{3} \AB{2}{1}{2} s_{23} \beta^2 } 
 \\ &
	+ \frac{2 m^3 \A{\eta_1}{\eta_4} \AB{3}{1}{2} \left( (\AB{2}{1}{2} + 2 m^2) \AB{2}{1}{2} + \AB{2}{1}{3} \AB{3}{1}{2} \right)}{\AB{2}{1}{2} \AB{2}{1}{3} s_{23} \beta^2 }
	+ \frac{2 m^3 \A{2}{\eta_1} \A{2}{\eta_4} \AB{3}{1}{2}^2}{\A{2}{3} \AB{2}{1}{3} s_{23} \beta^2 }
 \\ &
	- \frac{2 m^3 \A{2}{\eta_1} \A{3}{\eta_4} \AB{3}{1}{2} \left(2 \AB{2}{1}{2} + s_{23} \right)}{\A{2}{3} \AB{2}{1}{3} s_{23} \beta^2 } 
	- \frac{4 m \A{2}{\eta_1} \A{3}{\eta_4} \AB{3}{1}{2}^2 \left(\AB{2}{1}{2} + 2 m^2 \right)}{\A{2}{3} \AB{2}{1}{2} s_{23} \beta^2 } 
	\bigg) \\
& - (F_{2;23}^{m} - 2) 
    \frac{m \AB{3}{1}{2}^2 \left(\A{2}{\eta_4} \A{3}{\eta_1} + \A{2}{\eta_1} \A{3}{\eta_4} \right)}{\A{2}{3} s_{23}^2 \beta^2}
 - I_{3;2|3|41}^{m} \bigg(
	\frac{m \A{2}{\eta_1} \A{3}{\eta_4} \AB{3}{1}{2} \B{2}{3} }{\AB{2}{1}{3}}  \\ &
	- \frac{2 \A{\eta_1}{\eta_4} \AB{3}{1}{2}^2 m^3}{\AB{2}{1}{2} s_{23} \beta^2}
	+ \frac{2 m^3 \A{\eta_1}{\eta_4} \AB{3}{1}{2} \left(\AB{2}{1}{2} + 2 m^2 \right)}{\AB{2}{1}{3} s_{23} \beta^2}
    	- \frac{6 m^5 \A{3}{\eta_1} \A{3}{\eta_4} \AB{3}{1}{2} \B{2}{3}}{\AB{2}{1}{2}^2 s_{23} \beta^2} \\ &
	- \frac{m \A{2}{\eta_1} \A{3}{\eta_4} \AB{3}{1}{2} \left( 2 m^2 \AB{2}{1}{2}^2 - \AB{2}{1}{3} \AB{3}{1}{2} (\AB{2}{1}{2} + 4 m^2) \right)}{\A{2}{3} \AB{2}{1}{2}^2 \AB{2}{1}{3} \beta^2}
 \\ &
    	- \frac{m \A{2}{\eta_1} \AB{3}{1}{2} \left(\A{2}{\eta_4} \AB{3}{1}{2} s_{12} - \AB{\eta_4}{1}{2} \A{2}{3} \AB{2}{1}{2} \right)}{\A{2}{3} \AB{2}{1}{2} \AB{2}{1}{3} \beta^2}
	\bigg)
%
% \\ &
 + F_{2;12} \bigg(
	\frac{2 m \A{\eta_1}{\eta_4} \AB{3}{1}{2} s_{12}}{\AB{2}{1}{3} s_{23}}
\\ &
	- \frac{m^3 \A{3}{\eta_1} \A{3}{\eta_4} \AB{3}{1}{2}}{2 \A{2}{3} \AB{2}{1}{2} s_{12}}
	- \frac{m \A{2}{\eta_1} \A{2}{\eta_4} \AB{3}{1}{2}^2}{2 \A{2}{3} \AB{2}{1}{2} \AB{2}{1}{3}}
 	- \frac{m \A{2}{\eta_1} \A{3}{\eta_4} \AB{3}{1}{2}^2}{2 \A{2}{3} \AB{2}{1}{2} s_{12}}
	+ \frac{m \A{2}{\eta_1} \A{3}{\eta_4} \AB{3}{1}{2} }{\A{2}{3} \AB{2}{1}{3}}
	\bigg)
% \end{split}
% \end{align}
\\
% \begin{align}
% \begin{split}
	& - \left( F_{2;23}^{m} - 2\right) \times \bigg(
	\frac{m \A{\eta_1}{\eta_4} \AB{2}{1}{2} \AB{3}{1}{2}}{\AB{2}{1}{3} s_{23}} 
		+ \frac{m \A{\eta_1}{\eta_4} \AB{3}{1}{2} \left(\AB{2}{1}{2} + 2 m^2\right)}{\AB{2}{1}{3} s_{23} \beta^2} \\ &
		- \frac{m \AB{3}{1}{2} \left(2 \AB{2}{1}{2} + 3 s_{23} \right) \left(\A{3}{\eta_1} \A{3}{\eta_4} \AB{2}{1}{3} - \A{2}{\eta_1} \A{2}{\eta_4} \AB{3}{1}{2} \right)}{2 \A{2}{3} \AB{2}{1}{3} s_{23}^2 \beta^2} \\ &
		- \frac{m \A{2}{\eta_1} \A{3}{\eta_4} \AB{3}{1}{2} \left(2 \AB{2}{1}{2} + s_{23}\right)}{\A{2}{3} \AB{2}{1}{3} s_{23} \beta^2} 
	\bigg) 
+ \frac{m \AB{3}{1}{2}^2 \left( \A{\eta_1}{\eta_4} \AB{2}{1}{3} + \A{2}{\eta_1} \A{2}{\eta_4} \B{2}{3} \right)}{2 \AB{2}{1}{2} \AB{2}{1}{3} s_{23}}  \\ &
 	+ \frac{m \A{3}{\eta_4} \AB{3}{1}{2} \left(\A{3}{\eta_1} \AB{2}{1}{2} - \A{2}{\eta_1} \AB{3}{1}{2} \right)}{2 \A{2}{3} \AB{2}{1}{2} s_{12}}
 \end{split}
\end{align}

The primitive amplitudes amplitudes contributing to $A_{4;3}$ contain bubble, tadpole and rational 
terms that cancel when forming the complete amplitude. We do not list these terms explicitly in the 
following expressions.

\begin{align}
-i&\ampnorm A_4^{[L]} \left(1_t^{+},2^{+},4_\tb^{+},3^{+} \right)= \nonumber\\
& + I_{4;1|2|4|3} \times \left(
      \frac{m^3 \A{\eta_1}{\eta_4} \left(\AB{3}{1}{3}^2 - \AB{2}{1}{2} s_{23} \right)}{\A{2}{3}^2} 
      - \frac{m^3 \A{3}{\eta_1} \A{3}{\eta_4} \B{2}{3} \AB{2}{1}{2} \AB{3}{1}{3}}{2 \A{2}{3}^2 \AB{3}{1}{2}} \right) \nonumber\\ &
+  \left( \AB{2}{1}{2} \,  I_{3;12|3|4} + \AB{3}{1}{3} \, I_{3;13|2|4} \right) \times 
      \frac{m^3 \left(2 \A{\eta_1}{\eta_4} \AB{3}{1}{2} + \A{3}{\eta_1} \A{3}{\eta_4} \B{2}{3} \right)  }{2 \A{2}{3}^2 \AB{3}{1}{2}} \nonumber\\ &
+  \left( \AB{2}{1}{2} I_{3;12|3|4}^{m} + \AB{3}{1}{3} I_{3;13|2|4}^{m} \right) \times
    \frac{m^3 \left(2 \A{\eta_1}{\eta_4} \AB{3}{1}{2} - \A{3}{\eta_1} \A{3}{\eta_4} \B{2}{3} \right) }{2 \A{2}{3}^2 \AB{3}{1}{2}} \nonumber\\ &
+  \left( I_{3;12|3|4}^{m} - I_{3;13|2|4}^{m} \right) \times
    \frac{m^3 \B{2}{3} \left(\A{2}{\eta_1} \A{3}{\eta_4} + \A{3}{\eta_1} \A{2}{\eta_4} \right) }{\A{2}{3}^2} \nonumber\\ &
-  \left( \frac{I_{3;12|3|4}^{m}}{\AB{2}{1}{2}} + \frac{I_{3;13|2|4}^{m} }{\AB{3}{1}{3}} \right) \times
    \frac{m^3 \A{2}{\eta_1} \A{2}{\eta_4} \B{2}{3} \AB{3}{1}{2}}{\A{2}{3}^2} \nonumber\\ &
 + \text{ bubbles, tadpoles and rational terms  }
\end{align}

\begin{align}
-i&\ampnorm A_4^{[L]} \left(1_t^{+},2^{+},4_\tb^{+},3^{-} \right)= \nonumber\\
 & + I_{4;1|2|4|3} \times \bigg(
    \frac{m \left( \A{\eta_1}{\eta_4} \AB{3}{1}{2} - \A{3}{\eta_1} \A{3}{\eta_4} \B{2}{3} \right) \left(-\AB{2}{1}{3} \AB{3}{1}{2} + m^2 s_{23} \right)}{2 \AB{2}{1}{3}}  \nonumber\\ &
  + \frac{m^3 \AB{2}{1}{2} \AB{3}{1}{3} \left(\A{2}{\eta_1} \A{3}{\eta_4} \B{2}{3} + \A{\eta_1}{\eta_4} s_{12} \right)}{\AB{2}{1}{3}^2}
  - \frac{3 m^3 \A{3}{\eta_1} \A{3}{\eta_4} \AB{2}{1}{2} \AB{3}{1}{3}}{2 \A{2}{3} \AB{2}{1}{3}} \nonumber\\ &
  - \frac{2 m^3 \A{2}{\eta_1} \A{3}{\eta_4} \AB{2}{1}{2}^2 \AB{3}{1}{3}}{\A{2}{3} \AB{2}{1}{3}^2} 
  + \frac{m^3 \A{2}{\eta_1} \A{2}{\eta_4} \AB{2}{1}{2}^2 \AB{3}{1}{3}^2}{2 \A{2}{3} \AB{2}{1}{3}^3} 
 \bigg) \nonumber\\ &
 + \left( I_{3;12|3|4}^{m} - I_{3;12|3|4} \right) \times \bigg(
      \frac{m^3 \A{\eta_1}{\eta_4} \AB{2}{1}{2} s_{12}}{\AB{2}{1}{3}^2} 
     + \frac{m^3 \A{2}{\eta_1} \A{2}{\eta_4} \AB{2}{1}{2}^2 \AB{3}{1}{3}}{2 \A{2}{3} \AB{2}{1}{3}^3} \nonumber\\ &
     + \frac{m \AB{2}{1}{2} \left(\A{\eta_1}{\eta_4} \AB{3}{1}{2} - \A{3}{\eta_1} \A{3}{\eta_4} \B{2}{3} \right) }{2 \AB{2}{1}{3}}
     - \frac{3 m^3 \A{3}{\eta_1} \A{3}{\eta_4} \AB{2}{1}{2}}{2 \A{2}{3} \AB{2}{1}{3}} \nonumber\\ &
     - \frac{m^3 \A{2}{\eta_1} \A{3}{\eta_4} \AB{2}{1}{2} \B{2}{3}}{ \AB{2}{1}{3}^2}
     + \frac{2 m^3 \A{2}{\eta_1} \A{3}{\eta_4} \AB{2}{1}{2} \AB{3}{1}{3}}{\A{2}{3} \AB{2}{1}{3}^2}
    \bigg) \nonumber\\ 
& + \left( I_{3;13|2|4}^{m} - I_{3;13|2|4} \right) \times \bigg(
     \frac{m^3 \A{\eta_1}{\eta_4} \AB{3}{1}{3} s_{12} }{\AB{2}{1}{3}^2} 
     + \frac{m^3 \A{2}{\eta_1} \A{2}{\eta_4} \AB{2}{1}{2} \AB{3}{1}{3}^2}{2 \A{2}{3} \AB{2}{1}{3}^3} \nonumber\\ &
     + \frac{m \AB{3}{1}{3} \left(\A{\eta_1}{\eta_4} \AB{3}{1}{2} - \A{3}{\eta_1} \A{3}{\eta_4} \B{2}{3} \right) }{2 \AB{2}{1}{3}}
     - \frac{3 m^3 \A{3}{\eta_1} \A{3}{\eta_4} \AB{3}{1}{3}}{2 \A{2}{3} \AB{2}{1}{3}} \nonumber\\ &
     + \frac{m^3 \A{2}{\eta_1} \A{3}{\eta_4} \AB{3}{1}{3} \B{2}{3}}{\AB{2}{1}{3}^2}
     -  \frac{2 m^3 \A{2}{\eta_1} \A{3}{\eta_4} \AB{2}{1}{2} \AB{3}{1}{3}}{\A{2}{3} \AB{2}{1}{3}^2}
 \bigg) \nonumber\\& 
 + \text{ bubbles, tadpoles and rational terms  }
\end{align}

\subsection{Primitive amplitudes for $q\qb\to t\tb$}

\begin{align}
 \begin{split}
i&\ampnorm A_4^{[lc]}\left(1_t^{+},2_\qb^{+},3_q^{-},4_\tb^{+} \right) = \\
& \nnA_4^{(0)}\left(1_t^{+},2_\qb^{+},3_q^{-},4_\tb^{+} \right) \times \bigg\{
      \frac{s_{23} \AB{2}{1}{2}}{2} I_{4;1|2|3|4} - \frac{s_{23}}{2} I_{3;2|3|41}
      + \AB{2}{1}{2} I_{3;12|3|4} \\ &
      - \left( s_{23}  + \frac{3 m^2 (2 \AB{2}{1}{2} + s_{23})}{s_{23} \beta^4} \right) I_{3;1|23|4}
      - \frac{8}{3} I_{2;m} + \frac{23}{9} + \frac{12 m^2 (2 \AB{2}{1}{2} + s_{23}) }{s_{23}^2 \beta^4}  \\ &
      - \left(\frac{7}{6} + \frac{(2 \AB{2}{1}{2} + s_{23})}{2 s_{23} \beta^2} + \frac{6 m^2 (2 \AB{2}{1}{2} + s_{23})}{s_{23}^2 \beta^4} \right) \hat{I}_{2;23} 
      \bigg\} \\
& + \frac{m \left( \A{2}{\eta_1} \A{2}{\eta_4} \B{2}{3} + \A{\eta_1}{\eta_4} \AB{2}{1}{3} \right)}{2 \AB{2}{1}{3}^2} \times
      \bigg\{ \AB{2}{1}{2}^3 I_{4;1|2|3|4} - \AB{2}{1}{2}^2 I_{3;2|3|41} \\ &
	- \frac{2 \AB{2}{1}{2}^3}{s_{23}} I_{3;12|3|4} + \left(s_{12} s_{23} - 2 m^2 \AB{2}{1}{2} \right) I_{3;1|23|4}
      \bigg\} \\
& + \left(I_{3;1|23|4} + \frac{2}{s_{23}} ( \hat{I}_{2;23}- 2 ) \right) \times \bigg(
	  \frac{6 m^3 (\A{3}{\eta_1} \A{3}{\eta_4} \AB{2}{1}{2}^2 + \A{2}{\eta_1} \A{2}{\eta_4} \AB{3}{1}{2}^2) }{\A{2}{3} s_{23}^2 \beta^4} \\ &
	  - \frac{6 m^3 \A{2}{\eta_1} \A{3}{\eta_4} \AB{3}{1}{2} (2 \AB{2}{1}{2} + s_{23} ) }{\A{2}{3} s_{23}^2 \beta^4}
	  - \frac{3 m^3 \A{3}{\eta_1} \A{3}{\eta_4} }{2 \A{2}{3} \beta^4}
	\bigg) \\
& - \left( \hat{I}_{2;23} - 2 \right) \times \bigg(
       \frac{m \A{2}{\eta_1} \A{2}{\eta_4} \AB{3}{1}{2} \left( \AB{2}{1}{2} + 2 m^2 \right)}{\A{2}{3} \AB{2}{1}{3} s_{23} \beta^2} 
      + \frac{2 m \A{2}{\eta_1} \A{3}{\eta_4} \AB{3}{1}{2}}{\A{2}{3} s_{23} \beta^2} \\ &
      - \frac{m \A{3}{\eta_1} \A{3}{\eta_4} s_{12} }{\A{2}{3} s_{23} \beta^2}
      \bigg)  
 + I_{3;1|23|4} \bigg(
      \frac{m \left(\A{2}{\eta_1} \A{2}{\eta_4} \AB{3}{1}{2} + \A{3}{\eta_1} \A{3}{\eta_4} \AB{2}{1}{3} \right) \B{2}{3} }{2 \AB{2}{1}{3} } \\ &
      + \frac{m^3 \A{3}{\eta_1} \A{3}{\eta_4} \left(4 \AB{2}{1}{2} + s_{23} \right)}{2 \A{2}{3} s_{23} \beta^2} 
      - \frac{m^3 \A{2}{\eta_1} \A{2}{\eta_4} \AB{3}{1}{2} \left( 2 \AB{2}{1}{2} + s_{23} \right)}{\A{2}{3} \AB{2}{1}{3} s_{23} \beta^2}  \\ &
      - \frac{4 m^3 \A{2}{\eta_1} \A{3}{\eta_4} \AB{3}{1}{2}}{\A{2}{3} s_{23} \beta^2}
      \bigg)
 + \hat{I}_{2;23} \bigg(
      \frac{m (\A{3}{\eta_1} \A{3}{\eta_4} \AB{2}{1}{2}^2 + \A{2}{\eta_1} \A{2}{\eta_4} \AB{3}{1}{2}^2) }{\A{2}{3} s_{23}^2 \beta^2} \\ &
      - \frac{m \A{2}{\eta_1} \A{3}{\eta_4} \AB{3}{1}{2} (2 \AB{2}{1}{2} + s_{23}) }{\A{2}{3} s_{23}^2 \beta^2}
      - \frac{m^3 \A{3}{\eta_1} \A{3}{\eta_4} }{\A{2}{3} s_{23} \beta^2}
      \bigg)
 + F_{2;12} \bigg( 
    \frac{m \A{2}{\eta_1} \A{2}{\eta_4} \AB{3}{1}{2}^2}{s_{23} \A{2}{3} \AB{2}{1}{2}} \\ &
  + \frac{m \A{2}{\eta_1} \A{2}{\eta_4} \AB{2}{1}{2} \AB{3}{1}{2}}{s_{23} \A{2}{3} \AB{2}{1}{3}}
  - \frac{m \A{3}{\eta_4} (\A{2}{\eta_1} \AB{3}{1}{2} + m^2 \A{3}{\eta_1} ) }{\A{2}{3} \AB{2}{1}{2}}
 \bigg)\\
 \end{split} 
\end{align}

\begin{align}
-i&\ampnorm A_4^{[slc]} \left(1_t^{+},2_\qb^{+},3_q^{-},4_\tb^{+} \right)= \nn\\
& - \nnA_4^{(0)}\left(1_t^{+},2_\qb^{+},3_q^{-},4_\tb^{+} \right) \times \bigg( 
      s_{23} I_{3;2|3|41} + \left(s_{23} - 2 m^2 \right) I_{3;1|23|4}^{m} \nn\\ &
    + \frac{3}{2} \hat{I}_{2;23} + \left( 1 + \frac{2 \AB{2}{1}{2} + s_{23}}{2 s_{23} \beta^2} \right) F_{2;23}^{m} 
     + I_{2;m} - 4 - \frac{2 \AB{2}{1}{2} + s_{23}}{s_{23} \beta^2} \bigg) \nn\\
& - \left( F_{2;23}^{m} - 2 \right) \times \bigg(\frac{m^3 \A{3}{\eta_1} \A{3}{\eta_4}}{\A{2}{3} s_{23} \beta^2} 
   + \frac{m \A{2}{\eta_1} \A{3}{\eta_4} \AB{3}{1}{2} \left(2 \AB{2}{1}{2} + s_{23} \right)}{\A{2}{3} s_{23}^2 \beta^2} \nn\\ &
   - \frac{m \left(\A{3}{\eta_1} \A{3}{\eta_4} \AB{2}{1}{2}^2 + \A{2}{\eta_1} \A{2}{\eta_4} \AB{3}{1}{2}^2 \right)}{\A{2}{3} s_{23}^2 \beta^2} \bigg)  
\end{align}

\subsection{Fermion loop amplitudes}

Below we list the heavy fermion loop corrections for an arbitrary mass $m_H$. The light fermion loop 
contributions can be obtained by taking $m_H\to 0$.
\begin{align}
  -i&\ampnorm A_4^{[H]} \left(1_t^{+},2^{+},3^{+},4_\tb^{+} \right) = \nonumber\\
& - \frac{2 m ( \A{\eta_1}{\eta_4} \AB{2}{1}{2} - \A{2}{\eta_1} \A{3}{\eta_4} \B{2}{3})}{\A{2}{3}^3 \B{2}{3}} \times
    \bigg( s_{23} \, m_H^2 I_{3;2|3|41}^{m_H} 
      + 2 m_H^2 F_{2;23}^{m_H} + \frac{1}{6} s_{23} \bigg), \\ 
  -i& A_4^{[H]} \left(1_t^{+},2^{+},3^{-},4_\tb^{+} \right) = 0 \,,\\
  -i& A^{[H]}\left(1_t^+, 2_\qb^+, 3_q^-, 4_\tb^+ \right) = \nonumber\\&
  -\frac{2i}{3} A_4^{(0)}\left(1_t^+,2_\qb^+,3_q^-,4_\tb^+ \right) 
  \left(
  \left(2 \frac{m_H^2}{s_{23}} + 1 \right) \, F^{m_H}_{2;23} + I^{m_H}_{2;m} 
  -\frac{1}{3} \right) \, .
\end{align}

\section{Numerical Results \label{sec:num}}

The amplitudes in the previous section have been implemented into an efficient library for 
evaluation of the colour and helicity summed interference with tree level. In order to minimise the 
number of independent spinor products a specific choice of $\eta_1=\eta_4=p_2$ was made. This 
raises a number of issues since the symmetries between the helicity amplitudes are broken, however 
it is straightforward to generate all the necessary configurations automatically from those presented here.

For illustrative purposes we present numerical values for the unrenormalised amplitudes in the FDH scheme with the strong 
coupling set to one. We choose a generic phase-space point for the momenta as follows:
\begin{align}
  p_1 &= \left(-\sqrt{s+m^2},\sqrt{s},0,0\right), \\
	p_2 &= \sqrt{s+m^2}\left(1,\sin\theta,\cos\theta\cos\phi,\cos\theta\sin\phi\right), \\
	p_2 &= \sqrt{s+m^2}\left(1,-\sin\theta,-\cos\theta\cos\phi,-\cos\theta\sin\phi\right), \\
	p_4 &= \left(-\sqrt{s+m^2},-\sqrt{s},0,0\right). \\
\end{align}
The mass of the heavy quark is $m=1.75$. Numerical 
results at the point $s=1$, $\theta=\tfrac{\pi}{3}$, $\phi=\tfrac{\pi}{4}$, $N_f=5$, $N_H=1$, $N_c=3$ are given 
in Table~\ref{tab:numprimitive}. The interference with the tree level amplitude summed over helicity and 
colour is given in Table~\ref{tab:num} at two values of the renormalisation scale, $\mu_R^2=m^2$ and $\mu_R^2=4m^2$. 

\begin{table}
  \centering
  \begin{tabular}{|c|c|}
    \hline
    Primitive Amplitude & $\eps^0$ \\
\hline
\hline
$A_4^{(0)}(1_t^+,2^+,3^+,4_\tb^+)$ & $0.055220794+0.014807839i$\\
\hline
$A_4^{(0)}(1_t^+,2^+,3^-,4_\tb^+)$ & $0.062949503+0.14075936i$\\
\hline
\hline
$A_4^{[L]}(1_t^+,2^+,3^+,4_\tb^+)$ & $0.50453481+0.3385402i$\\
\hline
$A_4^{[R]}(1_t^+,2^+,3^+,4_\tb^+)$ & $0.021100789-0.12891563i$\\
\hline
$A_4^{[H]}(1_t^+,2^+,3^+,4_\tb^+)$ & $-0.01309239+0.028932428i$\\
\hline
$A_4^{[f]}(1_t^+,2^+,3^+,4_\tb^+)$ & $-0.0039170375-0.04534929i$\\
\hline
$A_{4;3}(1_t^+,2^+,3^+,4_\tb^+)$ & $0.63758829-1.1392369i$\\
\hline
$A_4^{[L]}(1_t^+,2^+,3^-,4_\tb^+)$ & $1.1550236+1.3088169i$\\
\hline
$A_4^{[R]}(1_t^+,2^+,3^-,4_\tb^+)$ & $2.5565516+1.0995254i$\\
\hline
$A_4^{[H]}(1_t^+,2^+,3^-,4_\tb^+)$ & $0$\\
\hline
$A_4^{[f]}(1_t^+,2^+,3^-,4_\tb^+)$ & $0$\\
\hline
$A_{4;3}(1_t^+,2^+,3^-,4_\tb^+)$ & $11.18323-1.475571i$\\
\hline
\hline
$A_4^{(0)}(1_t^+,2_\qb^+,3_q^-,4_\tb^+)$ & $0.85072714+0.25682619i$\\
\hline
\hline
$A_4^{[lc]}(1_t^+,2_\qb^+,3_q^-,4_\tb^+)$ & $8.0971525+3.2796876i$ \\
\hline
$A_4^{[slc]}(1_t^+,2_\qb^+,3_q^-,4_\tb^+)$ & $15.65914-4.9530347i$\\
\hline
$A_4^{[H]}(1_t^+,2_\qb^+,3_q^-,4_\tb^+)$ & $1.3699007+1.7416922i$\\
\hline
$A_4^{[f]}(1_t^+,2_\qb^+,3_q^-,4_\tb^+)$ & $0.24709861+2.0187408i$\\
\hline
  \end{tabular}
  \caption{Numerical values for the individual primitive amplitudes. The reference 
  vectors are chosen as $\eta_1=(3,2,2,1)$ and $\eta_4=(3,2,1,2)$, $\mu_R=2m$.}
  \label{tab:numprimitive}
\end{table}

\begin{table}
  \centering
  \begin{tabular}{|c|c|c|c|}
    \hline
    Channel & $\tfrac{1}{\eps^2}$ & $\tfrac{1}{\eps}$ & $\eps^0$ \\
    \hline
    $gg\to t\tb\,(\mu_R=m)$ & $-882.7183832$ & $1406.029038-1915.339983i$ & $2811.35321+1478.791625i$ \\
    \hline
    $gg\to t\tb\,(\mu_R=2m)$ & $-882.7183832$ & $182.3215209-1915.339983i$ & $3912.313922-1176.433394i$ \\
    \hline
    $q\qb\to t\tb\,(\mu_R=m)$ & $-82.70769231$ & $313.1028325+73.26833668i$ & $152.9616128-326.7068046i$ \\
    \hline
    $q\qb\to t\tb\,(\mu_R=2m)$ & $-82.70769231$ & $198.4456251+73.26833668i$ & $507.5399839-225.1353226i$ \\
    \hline
  \end{tabular}
  \caption{Numerical values for the interference between virtual and tree level amplitudes 
  summed over helicity and colour, $2\sum_{c,h}\mathcal{A}_4^{(1)}[\mathcal{A}_4^{(0)}]^*$.}
  \label{tab:num}
\end{table}

Our results have been cross-checked against previous calculations in the literature. At the 
amplitude level we find full agreement with numerical results of 
\cite{Ellis:ttggg}\footnote{A description of the reference vectors used to verify the results of 
ref. \cite{Ellis:ttggg} is given in Appendix~\ref{App:EGKMref}.}. We have also checked against the analytic 
results of ref. \cite{Anastasiou:ttgge2} up to $\mathcal{O}(\eps^0)$. We find full 
agreement with the implementation of the results of ref. \cite{Korner:ttgg} into {\tt MCFM}
\cite{mcfm}.

The FORTRAN program used to generate the results of Table~\ref{tab:num} evaluates the interference 
of one-loop amplitudes with the tree-level, summed over helicity and colour, in $43\mu$s for the 
gluon fusion channel and $13\mu$s for the quark annihilation channel. Roughly $60\%$ of this time is 
spent on the evaluation of the scalar integrals. The code was compiled using {\tt gfortran} with 
optimisation level {\tt-O2} and evaluated on a 2.93 GHZ Intel Core i3 530 CPU. The FORTRAN code {\tt 
BSYpptt} and the FORM files used to produce it are available from 
\url{http://www.nbia.dk/badger.html}. Both colour and helicity summed results and the individual 
primitive amplitudes are included.  

\section{Conclusions}

In this paper we have used the newly developed techniques of generalised unitarity to compute 
compact analytic representations of all helicity amplitudes relevant for heavy quark production at 
hadron colliders. Compact tree-level input was generated via BCFW recursion relations and the 
coefficients of the scalar integral computed via a purely algebraic procedure. A fully automated 
Feynman diagram approach was used to produce compact forms for the tadpole and rational terms.  

The calculation was performed in the spinor-helicity formalism with a completely general 
representation for the heavy quark spinors. The final amplitudes are expressed in terms 
of a relatively small set of spinor products. The analytic forms of the helicity amplitudes allow us 
to take a new look at the structure of the one-loop helicity amplitudes. The all-plus helicity 
configuration in the gluon channel takes a remarkably simple structure. In comparison to the well 
known MHV structure in massless amplitudes it is expected that such a simplicity persists at higher 
multiplicity. The most notable feature of this new representation is the cancellation of bubbles and rational 
terms in the sub-leading colour contribution to $gg\to t\tb$ which is expected to be related to the 
stronger UV constraints on this sector. Together with similar features found in other analytic 
computations \cite{Badger:2009vh}, this may motivate future investigations.

The final results yield a flexible implementation suitable for computations of spin correlations and 
decays in the narrow width approximation.

We have demonstrated that such techniques provide a feasible method to calculate analytic one-loop 
amplitudes with full mass dependence and serve as a solid base for future implementations of higher 
multiplicity amplitudes.

\begin{acknowledgments}
We are particularly grateful to Keith Ellis, John Campbell, Peter Uwer and Sven-Olaf Moch for helpful discussions and 
feedback throughout this project. The work of VY was supported by DFG SFB-TR-9 and by LHCPhenoNet 
network PITN-GA-2010-264564. The work of SB has been supported in part by Danish Natural 
Science Research Council grant 10-084954. The work of RS was supported in part by grant GK 1504 of 
the DFG Graduiertenkolleg.
\end{acknowledgments}

\appendix

\section{Tree-level amplitudes \label{App:trees}}

For completeness we present the tree-level amplitudes relevant for our computation using the spinor-helicity formalism described
in section \ref{sec:spinors}.

\subsection{On-shell Three-point Vertices}

The independent on-shell three-point vertices are:
\begin{align}
	-iA_3(1_t^+,2^+,3_\tb^+) &=
	\frac{\AB{\xi}{1}{2}}{\A{\xi}{2}}
	\frac{ m \A{\eta_1}{\eta_3}}{\A{\eta_1}{\fl 1}\A{\eta_3}{\fl 3}},
	\label{eq:ppp}\\
	-iA_3(1_t^-,2^+,3_\tb^-) &= 
	\frac{\AB{\xi}{1}{2}}{\A{\xi}{2}}
	\frac{\A{\fl1}{\fl3}}{m},
	\label{eq:mpm}
\end{align}
\begin{align}
	-iA_3(1_t^-,2^+,3_\tb^+) &= -
	\frac{\AB{\xi}{1}{2}}{\A{\xi}{2}}
	\frac{\A{\fl1}{\eta_3}}{\A{\fl 3}{\eta_1}},
	\label{eq:mpp}\\
	-iA_3(1_t^+,2^+,3_\tb^-) &=
	\frac{\AB{\xi}{1}{2}}{\A{\xi}{2}}
	\frac{\A{\fl3}{\eta_1}}{\A{\fl 1}{\eta_3}}.
	\label{eq:ppm}
\end{align}

\subsection{$gg\to t\tb$ Tree Amplitudes}
For the $gg\to\tb t$ channel with adjacent fermions we obtain 
\begin{align}
  -i A^{(0)}_4(1_t^+,2^+,3^+,4_\tb^+) &=
    \frac{m^3 \A{\eta_1}{\eta_4} \B{2}{3}}{\A{2}{3}\AB{2}{1}{2}\A{\eta_1}{\fl 1}\A{\eta_4}{\fl 4}} \,,\\
  -i A^{(0)}_4(1_t^+,2^+,3^-,4_\tb^+) &=	
  \frac{m \AB{3}{1}{2} \left( \A{\eta_1}{\eta_4}\AB{3}{1}{2} - \B{2}{3} \A{3}{\eta_1} \A{3}{\eta_4} \right) }{s_{23} \AB{2}{1}{2} \A{\eta_1}{\fl 1}\A{\eta_4}{\fl 4}} \, .
\end{align}
For the sub-leading colour contributions we also make use of compact forms for the case of
non-adjacent fermions:
\begin{align}
  -i A^{(0)}_4(1_t^+,2^+,4_\tb^+,3^+) &=
    \frac{m^3 \A{\eta_1}{\eta_4} \B{2}{3}^2}{\A{\eta_1}{\fl 1} \A{\eta_4}{\fl 4} \AB{2}{1}{2} \AB{3}{1}{3} },\\
  -i A^{(0)}_4(1_t^+,2^+,4_\tb^+,3^-) &=-
    \frac{m \AB{3}{1}{2} (\A{\eta_1}{\eta_4} \AB{3}{1}{2} - \B 23 \A{3}{\eta_1} \A{3}{\eta_4} 
    )}{\A{\eta_1}{\fl 1} \A{\eta_4}{\fl 4} \AB{2}{1}{2} \AB{3}{1}{3}}.	
\end{align}
The other fermion helicity states can be obtained via the relation given in \eqref{eq:helswapid}.

\subsection{$q\qb\to t\tb$ Tree Amplitudes}

There is only one independent helicity amplitude in this channel which can be written as,
\begin{align}
  -iA^{(0)}_4(1_t^+,2^+_\qb,3^-_q,4_\tb^+) = 
    \Frac[m\left(\A{\eta_1}{3}\AB{\eta_4}{4}{2} + \A{\eta_4}{3}\AB{\eta_1}{1}{2}\right),\A{\eta_1}{\fl1}\A{\eta_4}{\fl4}s_{23}] \, . 
\end{align}

\section{Conversion to Four Component Dirac Spinors \label{App:EGKMref}}

We note that the formalism of equations (\ref{eq:Usp_m}) and (\ref{eq:Vsp_m}) can be connected with a more conventional approach to massive
solutions of the Dirac equation by choosing a specific reference frame. For a massive four-vector
$Q^\mu=(E,Q_1,Q_2,Q_3)$ with $Q^2=m^2$ we first define:
\begin{align}
	Q_+ &= E+Q_3, & Q_- &= E-Q_3, \\
	Q^\perp &= Q_1+iQ_2, & \bar{Q}^\perp &= Q_1-iQ_2.
\end{align}
Making a choice of
\begin{equation}
	\eta = \frac{1}{2(Q_-+m)^2}
	\begin{pmatrix}
		(Q_-+m)^2+Q_1^2+Q_2^2\\
		2Q_1(Q_-+m)\\
		2Q_2(Q_-+m)\\
		-(Q_-+m)^2+Q_1^2+Q_2^2\\	
	\end{pmatrix},
\end{equation}
then yields the following four dimensional representations:
\begin{align}
	u_+(Q,m)&=\begin{pmatrix} \sqrt{E+m} \\ 0 \\ \frac{Q_3}{\sqrt{E+m}} \\ 
	  \frac{Q^\perp}{\sqrt{E+m}} \end{pmatrix},
		&
	u_-(Q,m)&=\begin{pmatrix} 0 \\ \sqrt{E+m}\\ \frac{\bar{Q}^\perp}{\sqrt{E+m}} \\ 
	  -\frac{Q_3}{\sqrt{E+m}} \end{pmatrix},
		\\
	v_+(Q,m)&=\begin{pmatrix} \frac{Q_3}{\sqrt{E+m}} \\ \frac{Q^\perp}{\sqrt{E+m}} \\ \sqrt{E+m} 
	  \\ 0 \end{pmatrix},
		&
	v_-(Q,m)&=\begin{pmatrix} \frac{\bar{Q}^\perp}{\sqrt{E+m}} \\ -\frac{Q_3}{\sqrt{E+m}} \\ 0 
	  \\ \sqrt{E+m} \end{pmatrix}.
\end{align}
This allows a simple way to compare analytic results with numerical ones in the literature.

\providecommand{\href}[2]{#2}
\begingroup\raggedright\endgroup

\end{document}